\documentclass[preprint,nopreprintline]{elsarticle}

\usepackage{lineno,hyperref}
\usepackage{xcolor}

\usepackage{amsmath,amssymb,amsthm,amsfonts,amscd,amsbsy,amsxtra,bm}
\usepackage{mathtools}
\usepackage{subcaption}
\usepackage{longtable}

\usepackage{booktabs}
\usepackage{etoolbox}

\hypersetup{
    hidelinks,
    pdfauthor=Lukas Petrich; Georg Lohrmann; Fabio Martin; Albert Stoll; Volker Schmidt
}

\biboptions{authoryear}

\usepackage{enumitem}
\setlistdepth{9}

\setlist[itemize,1]{label=$\bullet$}
\setlist[itemize,2]{label=$\bullet$}
\setlist[itemize,3]{label=$\bullet$}
\setlist[itemize,4]{label=$\bullet$}
\setlist[itemize,5]{label=$\bullet$}
\setlist[itemize,6]{label=$\bullet$}
\setlist[itemize,7]{label=$\bullet$}
\setlist[itemize,8]{label=$\bullet$}
\setlist[itemize,9]{label=$\bullet$}
\renewlist{itemize}{itemize}{9}

\newcommand{\R}{\mathbb{R}}

\newcommand{\meter}{\ensuremath{\mathrm{m}}}
\newcommand{\centimeter}{\ensuremath{\mathrm{cm}}}
\newcommand{\millimeter}{\ensuremath{\mathrm{mm}}}
\newcommand{\squareMeter}{\ensuremath{\mathrm{m}\textsuperscript{\tiny 2}}}

\newcommand{\strategyLoneRobot}{\textsc{Robot}}
\newcommand{\strategyLoneFarmer}{\strategyLoneRobot{}}
\newcommand{\strategySnakingTractor}{\textsc{Tractor}}

\newcommand{\strategyParamActionThreshold}{\ensuremath{\ell_a}}
\newcommand{\strategyParamMeanderingWidthInM}{\ensuremath{w_\mathrm{m}}}
\newcommand{\strategyParamDilationRadius}{\ensuremath{r_\mathrm{t}}}
\newcommand{\strategyParamTreatmentRadius}{\strategyParamDilationRadius}
\newcommand{\strategyParamDefaultTreatmentLength}{\ensuremath{l_\mathrm{d}}}

\newcommand{\strategyParamToolSegmentCount}{\ensuremath{n_\mathrm{s}}}

\newcommand{\strategyStartingPoint}{x_0}
\newcommand{\strategyInvestedArea}{M_{\mathrm{inf}}}
\newcommand{\strategyTractorPrimaryDrivingDirection}{u_\mathrm{d}}

\newcommand{\TreatedIdentifier}{t}
\newcommand{\resTreatedArea}{\ensuremath{M_\mathrm{\TreatedIdentifier}}}
\newcommand{\resTreatedPattern}{s^{\mathrm{(\TreatedIdentifier)}}}
\newcommand{\resTreatedPointsNum}{n^{\mathrm{(\TreatedIdentifier)}}}

\newcommand{\metricHighestDensityGrid}{G}

\newcommand{\metricDrivingDistanceInM}{\ensuremath{d_\mathrm{d}}}
\newcommand{\metricTreatedAreaFrac}{\ensuremath{A_\mathrm{t}}}

\newcommand{\metricFracRemainingWeeds}{\ensuremath{f_\mathrm{r}}}
\newcommand{\metricHighestRemainingWeedsDensityM}[1]{\ensuremath{\rho_\mathrm{r}}}
\newcommand{\metricTreatedAreaPerWeed}{\ensuremath{A_{\mathrm{eff}}}}

\newcommand{\datasetEarlyIdentifier}{1}
\newcommand{\datasetLateIdentifier}{2}
\newcommand{\datasetHalfIdentifier}{<}
\newcommand{\datasetStandardIdentifier}{0}
\newcommand{\datasetDoubleIdentifier}{>}

\newcommand{\datasetExperimentalData}{\textsc{Exp}}
\newcommand{\datasetExperimentalDataEarlyStandard}{\datasetExperimentalData{}\textsuperscript{\datasetEarlyIdentifier{}}}
\newcommand{\datasetExperimentalDataLateStandard}{\datasetExperimentalData{}\textsuperscript{\datasetLateIdentifier{}}}

\newcommand{\datasetCalibratedInhomogeneousPoisson}{\textsc{Cal}}
\newcommand{\datasetCalibratedInhomogeneousPoissonEarlyStandard}{\datasetCalibratedInhomogeneousPoisson{}$^{\datasetEarlyIdentifier{}}_{\datasetStandardIdentifier{}}$}
\newcommand{\datasetCalibratedInhomogeneousPoissonEarlyDouble}{\datasetCalibratedInhomogeneousPoisson{}$^{\datasetEarlyIdentifier{}}_{\datasetDoubleIdentifier{}}$}
\newcommand{\datasetCalibratedInhomogeneousPoissonEarlyHalf}{\datasetCalibratedInhomogeneousPoisson{}$^{\datasetEarlyIdentifier{}}_{\datasetHalfIdentifier{}}$}

\newcommand{\datasetSyntheticCenteredPoisson}{\textsc{Cen}}
\newcommand{\datasetSyntheticCenteredPoissonEarlyStandard}{\datasetSyntheticCenteredPoisson{}$^{\datasetEarlyIdentifier{}}_{\datasetStandardIdentifier{}}$}

\newcommand{\datasetSyntheticHomogeneousPoisson}{\textsc{Hom}}
\newcommand{\datasetSyntheticHomogeneousPoissonEarlyStandard}{\datasetSyntheticHomogeneousPoisson{}$^{\datasetEarlyIdentifier{}}_{\datasetStandardIdentifier{}}$}

\newcommand{\datasetSyntheticSinusoidalPoisson}{\textsc{Sin}}

\newcommand{\GroundTruthIdentifier}{gt}
\newcommand{\ObservationIdentifier}{obs}
\newcommand{\pppSimWindow}{\mathcal{W}}
\newcommand{\pppGTProcess}{S^{\mathrm{(\GroundTruthIdentifier)}}}

\newcommand{\pppGTPattern}{s^{\mathrm{(\GroundTruthIdentifier)}}}   
\newcommand{\pppGTPointsNum}{n^{\mathrm{(\GroundTruthIdentifier)}}}
\newcommand{\pppCountingMeasure}[1]{\Phi\left(#1\right)}
\newcommand{\pppIntensityFunction}{\lambda}
\newcommand{\pppBaseIntensityFunction}{\pppIntensityFunction_0}
\newcommand{\pppIntensityMeasure}[1]{\Lambda\left(#1\right)}
\newcommand{\pppIntensityFunctionNormalization}{\alpha}
\newcommand{\pppIntensityFactor}{\lambda^*}

\newcommand{\pppObservedPattern}{s^{\mathrm{(\ObservationIdentifier)}}}   
\newcommand{\pppObservedPointsNum}{n^{\mathrm{(\ObservationIdentifier)}}}

\newcommand{\pppObservationThinningProbEarly}{p^{\mathrm{(\ObservationIdentifier)}}_{\datasetEarlyIdentifier}}
\newcommand{\pppObservationThinningProbLate}{p^{\mathrm{(\ObservationIdentifier)}}_{\datasetLateIdentifier}}
\newcommand{\TargetIdentifier}{tgt}
\newcommand{\pppTargetedPattern}{s^{\mathrm{(\TargetIdentifier)}}}   
\newcommand{\pppTargetedPointsNum}{n^{\mathrm{(\TargetIdentifier)}}}   

\newcommand{\pppExpPattern}{s^{\mathrm{(\datasetExperimentalData)}}}
\newcommand{\pppExpPointsNum}{n^{\mathrm{(\datasetExperimentalData)}}}
\newcommand{\pppExpEarlyPointsNum}{n^{\mathrm{(\datasetExperimentalData)}}_{\mathrm{\datasetEarlyIdentifier}}}
\newcommand{\pppExpLatePointsNum}{n^{\mathrm{(\datasetExperimentalData)}}_{\mathrm{\datasetLateIdentifier}}}

\newcommand{\pppIntFunCalibratedKernel}{k^{\mathrm{(\datasetCalibratedInhomogeneousPoisson)}}}
\newcommand{\pppIntFunCalibratedBandwidth}{h^{\mathrm{(\datasetCalibratedInhomogeneousPoisson)}}}
\newcommand{\pppIntFunCenteredMean}{\mu^{\mathrm{(\datasetSyntheticCenteredPoisson)}}}
\newcommand{\pppIntFunCenteredCov}{\Sigma^{\mathrm{(\datasetSyntheticCenteredPoisson)}}}
\newcommand{\pppIntFunSinDirection}{u^{\mathrm{(\datasetSyntheticSinusoidalPoisson)}}}
\newcommand{\pppIntFunSinWaveLength}{\lambda^{\mathrm{(\datasetSyntheticSinusoidalPoisson)}}}

\renewcommand{\Pr}{\mathbb{P}}

\begin{document}

\begin{frontmatter}

\title{Model-based scenario analysis for effective site-specific weed control on grassland sites}

\author[uulmaddress]{Lukas Petrich\corref{corauth}}
\cortext[corauth]{Corresponding author}
\ead{lukas.petrich@uni-ulm.de}

\author[hfwuadress]{Georg Lohrmann}
\author[hfwuadress]{Fabio Martin}
\author[hfwuadress]{Albert Stoll}
\author[uulmaddress]{Volker~Schmidt}

\address[uulmaddress]{Institute of Stochastics, Ulm University, D-89069 Ulm, Germany}
\address[hfwuadress]{Hochschule für Wirtschaft und Umwelt Nürtingen-Geislingen, Neckarsteige 6-10,\\D-72622 Nürtingen, Germany}

\begin{abstract} 
The site-specific management of weeds in grassland is often challenging because different weed control strategies have different trade-offs regarding the required resources and treatment efficiency. So, the question arises whether a wide tractor-based system with section control or a small agricultural robot has a higher weed control performance for a given infestation scenario. For example, a small autonomous robot moving from one weed to the next might have much shorter travel distances (and thus lower energy and time costs) than a tractor-mounted system if the locations of the weeds are relatively isolated across the field. However, if the plants are highly concentrated in small areas so-called clusters, the increased width of the tractor-mounted implement could be beneficial because of shorter travel distances and greater working width.

An additional challenge is the fact that there is no complete knowledge of the weed locations. Weeds may not have been detected, for example, due to their growth stage, occlusion by other objects, or misclassification. Weed control strategies must therefore also be evaluated with regard to this issue. Thus, in addition to the driving distance, other metrics are also of interest, such as the number of plants that were actually controlled or the size of the total treatment area.

We performed this investigation for the treatment of the toxic \textit{Colchicum autumnale}, which had been detected in drone images of extensive grassland sites. In addition to real data, we generated and analyzed simulated weed locations using mathematical models of stochastic geometry. These offer the possibility to simulate very different spatial distributions of toxic plant locations. Different treatment strategies were then virtually tested on this data using Monte Carlo simulations and their performance was statistically evaluated.
\end{abstract}


\end{frontmatter}


\section{Introduction}

For the management of weeds on grassland sites, different treatment tools and strategies are available. In order to keep the costs at a minimum and to reduce the impact on the surrounding flora and fauna, which might be required by environment protection regulation, site-specific weed control has proven preferable (see, e.g., \cite{Schellberg2008,Wegener2020}). With these techniques, however, new questions arise such as how a treatment strategy fares with different spatial distributions and severities of weeds on a field. In addition to this, site-specific methods rely on precise locations of the targeted weeds. Regardless of how these locations are acquired, a complete and perfect survey of the weed population is practically infeasible. It is therefore also important to investigate how treatments based on partial information perform with respect to the whole (unobserved) weed population.

For these reasons we performed a virtual scenario analysis to study different treatment strategies. Simulated weed locations allowed to create several prototypical infestation scenarios with varying severities. Additionally, experimental observations of \textit{Colchicum autumnale} plants on an extensive grassland site were also employed to provide a realistic baseline for the simulation study. Here, \textit{C. autumnale} are toxic plants, which pose a thread to farm animals especially in hay or silage. Note, however, that the present paper does not present a thorough stochastic model for the experimental \textit{C. autumnale} locations, which would require much more empirical data over a longer timespan. With the proposed simulation framework it is possible to illuminate the aforementioned questions in silico without the need for extensive experimental setup (e.g., building of the treatment tool, finding suitable fields, etc.). The considered treatment tools are inspired by the non-chemical weed control tools developed in \citet{Stoll2020} and \citet{Martin2022}---that is a small autonomous robot with circular mower and a water-hydraulic tractor-based system with section control---, but the results apply similarly to other tools such as site-specific herbicide sprayers, or spot-spraying through unmanned aerial vehicles (UAV). Here, the two considered treatment tools differ primarily in terms of flexibility---the autonomous robot is able to move from one targeted location to the next, while the tractor traverses the field in meandering lines---and working width.  One of the main goals of the present paper is thus to evaluate in which infestation scenarios one treatment tool outperforms the other especially in the light of unobserved weed locations.

Simulation studies are certainly not new when investigating weeds and their control strategies, see for example \citet{Somerville2020} for an overview of different weed models, or \citet{Colbach2018} for a study of landsharing versus landsparing. However, to the best of our knowledge this is the first paper that deals with site-specific treatment tools on grassland.

\section{Materials and methods}
The present paper deals with two kinds of weed locations---experimentally observed and simulated ones. In this section, we first describe the acquisition and preprocessing of the experimental dataset. After that, the methods for simulating weed locations and the virtual weed control tools are laid out.

\subsection{Data acquisition}
The experimental dataset consists of locations of \textit{C. autumnale} flowers that have been extracted from drone images. The images were acquired on September 3rd and 10th, 2019, on an extensive grassland field near Nürtingen, Germany. The field had an area of
about 8256\,\squareMeter{}, see Figure~\ref{fig:experimentalData}.
At the two observation dates different \textit{C. autumnale} plants were blooming leading to varying numbers of visible flowers. The camera, a Sony alpha 7 RII with a CMOS full-frame 42.4 MP image sensor and its lens with 24\,\millimeter{} focal length, was mounted on a HiSystems MK ARF-OktoXL 4S12 octocopter. The route of the drone was about 10\,\meter{} above ground and was chosen such that the resulting images had an overlap of about 55\%.

The individual drone images where stitched to two orthomosaics (one for each observation date) using the Agisoft Metashape software. In the same step, the orthomosaics were georeferenced by matching markers (ground control points) that were placed on the field and whose GPS positions were obtained by real-time kinematic positioning with their corresponding pixel coordinates.

\subsection{Data preprocessing}
\subsubsection{Image registration}

In addition to the georeferencing, the overlap between the two orthomosaics has been further improved by pixel-based image registration. For this, matching pairs of control points in the two images have been created by visual inspection at objects such has fence posts or trees that remained unchanged between the two observations. Then, MathWorks MATLAB was used to find a coordinate transformation that transforms one orthomosaic to match the second one.  Here, the `projective' transformation type \citep{Jaehne2005, matlab_image} was chosen out of `nonreflective', `similarity', `affine', `projective', and `polynomial' with degrees up to 4 (see \citet{matlab_image}) as a trade-off between minimizing the Euclidean distance between the first set of control points and the transformed control points of the second orthomosaic and visual goodness-of-fit. For both aligned orthomosaics a region of interest $\pppSimWindow \subset \R^2$ was defined by removing all parts of the images that did not show the considered field or where information from one observation was missing.  In the following, only this cutout of the orthomosaics is considered.

\subsubsection{Observed weed locations}\label{sec:ObservedExpLocations}
For the two aligned orthomosaics, \textit{C. autumnale} flowers were automatically detected using the predictor developed in \cite{Petrich2020}. Manual checking of each predicted weed location and visual inspection of the remaining images ensured an accurate survey of weed locations at the two observation dates. Here, we considered locations that are closer than 5\,\centimeter{} to each other to be the same weed and replaced these locations with their centroid.

\begin{figure}
    \begin{subfigure}[t]{0.3\linewidth}
        \centering
        \includegraphics[width=\linewidth]{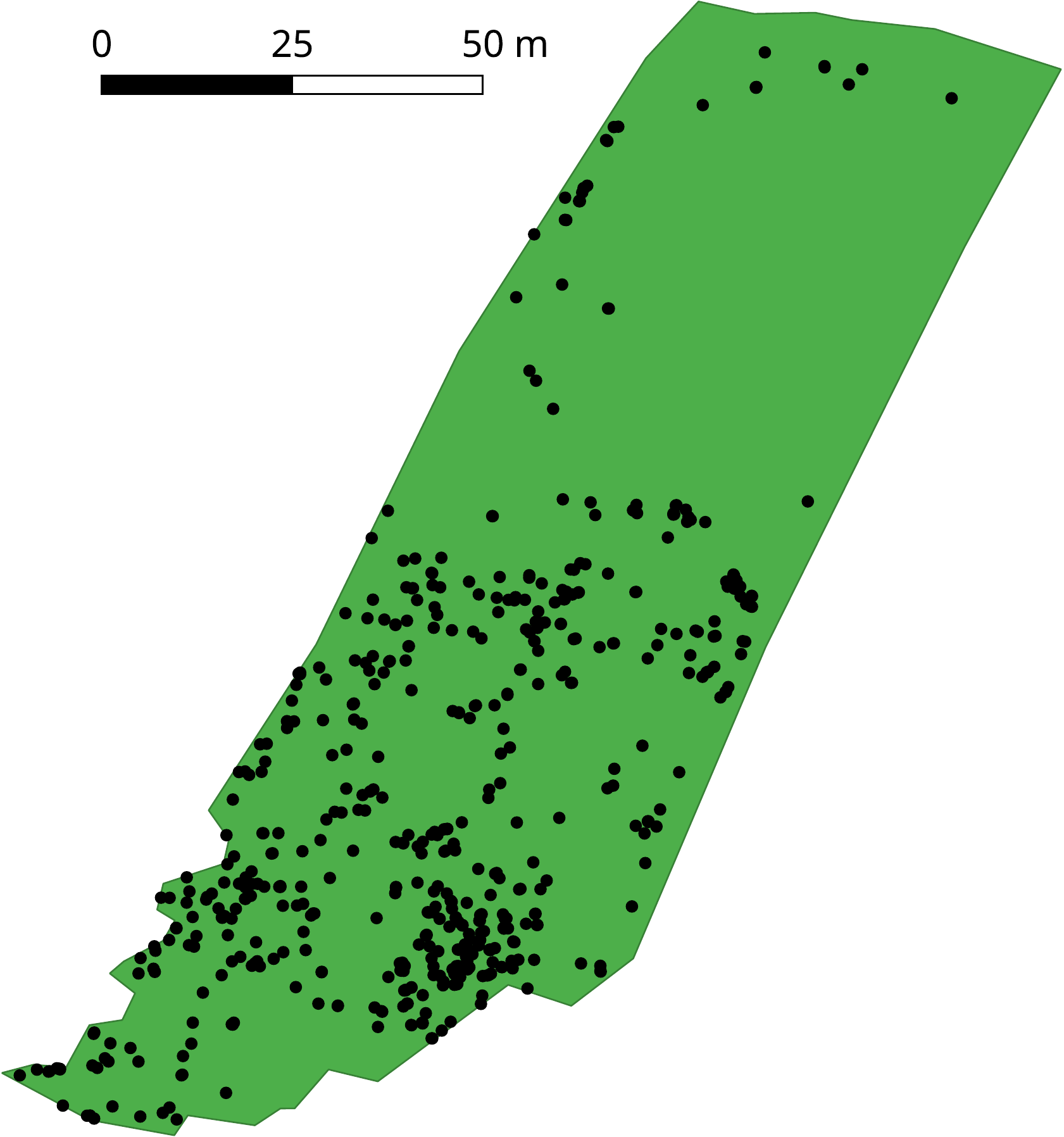}
        \caption{Experimentally observed weed locations from the 3rd September, 2019 (\datasetExperimentalDataEarlyStandard{}).}
        \label{fig:expDataObservationEarly}
    \end{subfigure}\hfill
    \begin{subfigure}[t]{0.3\linewidth}
        \centering
        \includegraphics[width=\linewidth]{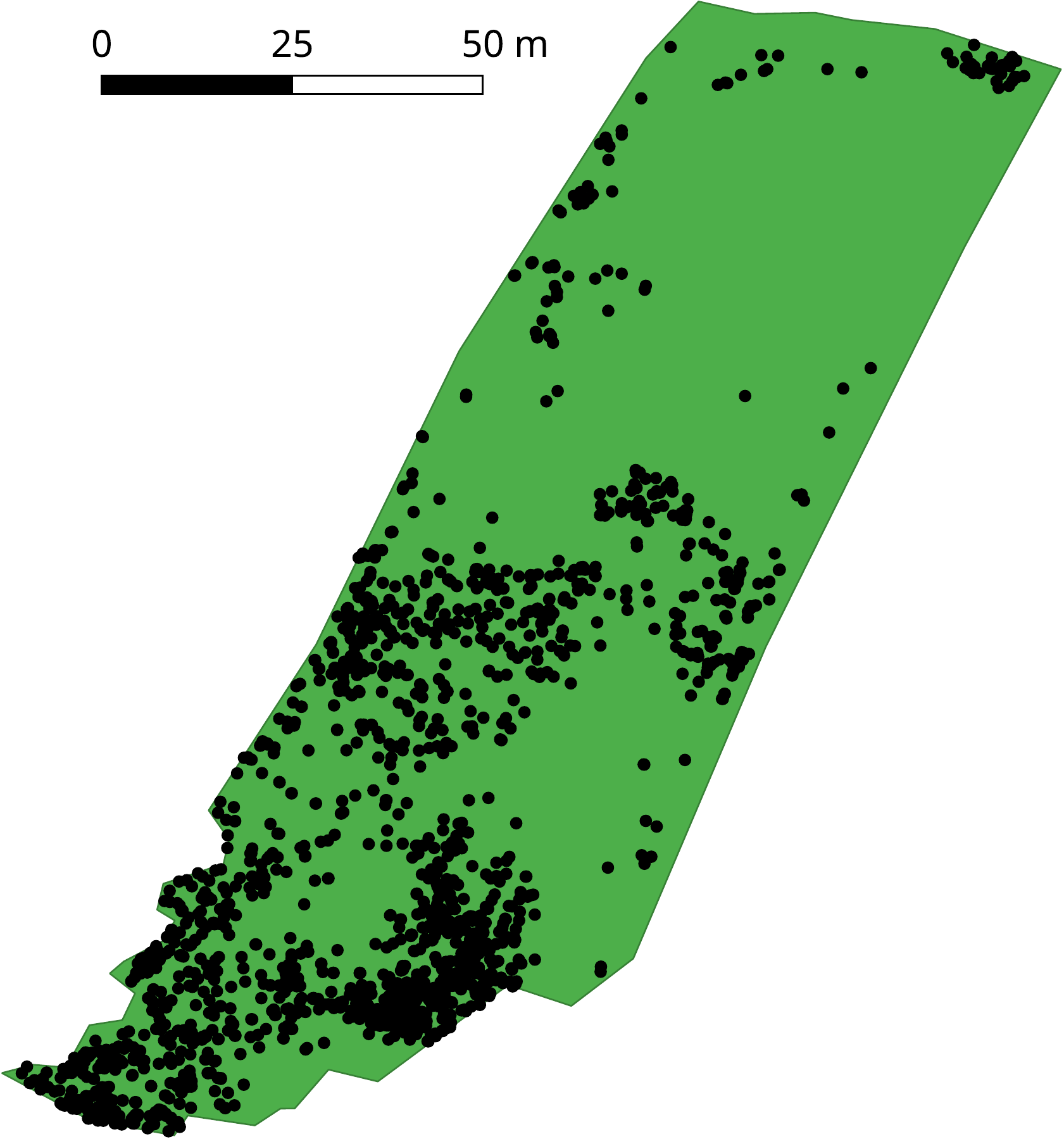}
        \caption{Experimentally observed weed locations from the 10th September, 2019 (\datasetExperimentalDataLateStandard{}).}
        \label{fig:expDataObservationLate}
    \end{subfigure}\hfill
    \begin{subfigure}[t]{0.3\linewidth}
        \centering
        \includegraphics[width=\linewidth]{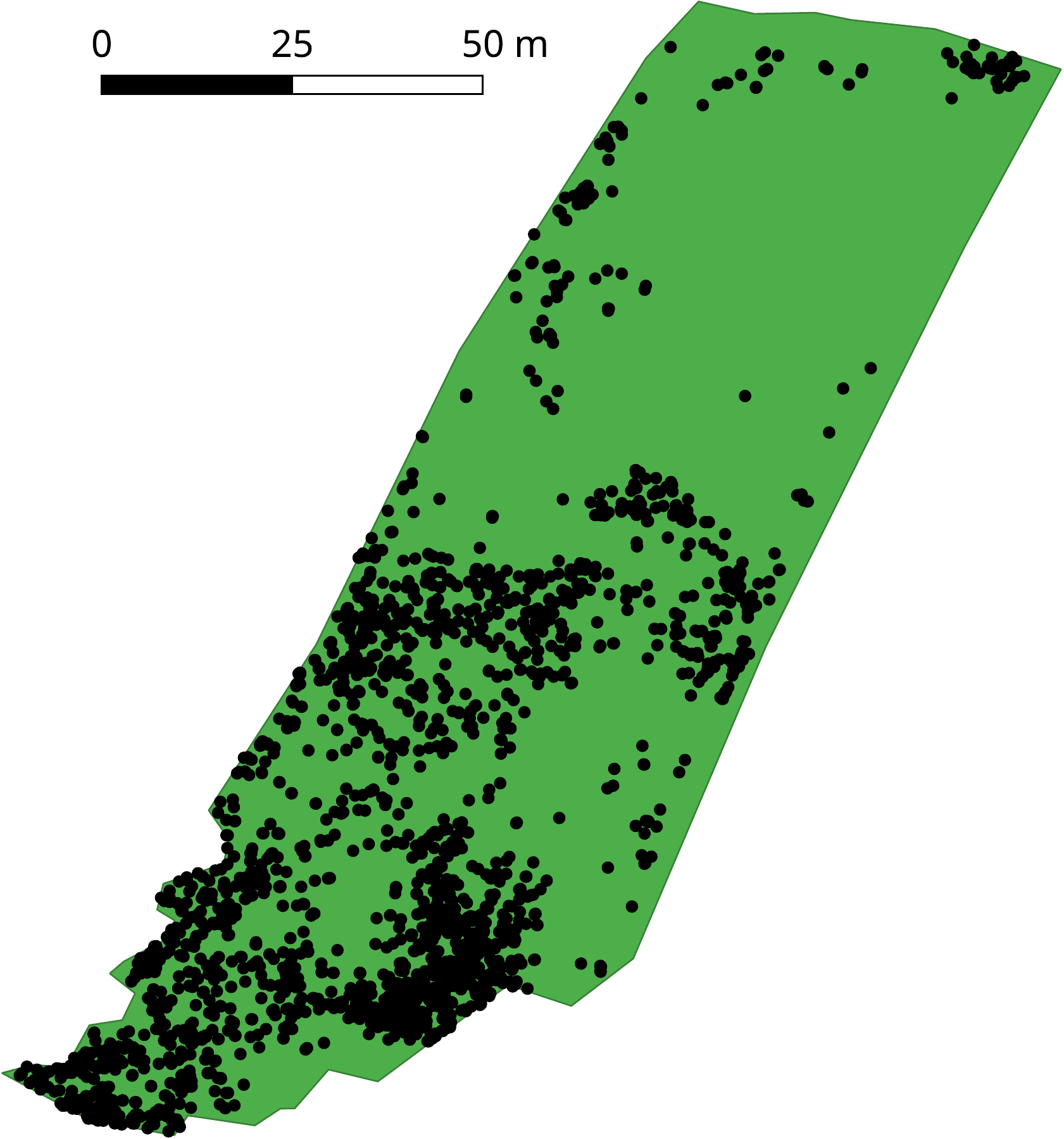}
        \caption{Combination of the observed weed locations \datasetExperimentalDataEarlyStandard{} and \datasetExperimentalDataLateStandard{} acting as the ground truth locations \datasetExperimentalData{}.}
        \label{fig:expDataMerged}
    \end{subfigure}
    \caption{Visualization of  experimental data. The green area corresponds to the whole field.}\label{fig:experimentalData}
\end{figure}

The resulting weed locations are shown in Figures~\ref{fig:expDataObservationEarly} and \ref{fig:expDataObservationLate}. In total there were $\pppExpEarlyPointsNum = 550$ detected flowers in the observation from September 3rd, 2019 and $\pppExpLatePointsNum = 1792$ flowers from September 10th. In the following we refer to the first as \datasetExperimentalDataEarlyStandard{} and to the latter as \datasetExperimentalDataLateStandard{}.

\subsubsection{Experimental ground truth weed locations}\label{sec:GroundTruthExpLocations}
Obviously, the two datasets \datasetExperimentalDataEarlyStandard{} and \datasetExperimentalDataLateStandard{} were partial observations of the complete weed population on the field. This incompleteness of the observations could have multiple reasons such as the plants were not yet in a growth state in which they can be detected, they were occluded by other objects, etc. For the present simulation study, we made the simplifying assumption that the locations of the whole population can be obtained by combining the observed datasets \datasetExperimentalDataEarlyStandard{} and \datasetExperimentalDataLateStandard{}, see Figure~\ref{fig:expDataMerged}.  This is possible since we do not aim to provide an accurate modeling of the true weed population, but rather only use the experimental data as a rough baseline for our scenario analysis. Note, however, that for the combination, we again considered locations closer than 5\,\centimeter{} to each other to be the same weed and replaced these locations with their centroid. The resulting experimental ground truth weed location dataset, denoted \datasetExperimentalData{} in the following, comprised a total of $\pppExpPointsNum =  2313$ locations. So in summary, we have a ground truth dataset (\datasetExperimentalData{}) comprised of all weed locations and two subsets thereof (\datasetExperimentalDataEarlyStandard{} and \datasetExperimentalDataLateStandard{}) corresponding to the weeds that were observed at the two observation dates (September 3rd and 10th, respectively). This general setup will be maintained for the simulated data, see Section~\ref{sec:simulatedData}.

\subsection{Simulated data}\label{sec:simulatedData}
In addition to real experimental data, we also considered simulated data, i.e., we generated and analyzed simulated weed locations using mathematical models of stochastic geometry, see \citet{Chiu2013}.
This allowed us to investigate various scenarios not readily available with experimental data. The general setup was to first simulate ground truth weed locations. However, in practice only a subset of the actual weed locations is observed. We imitate this partial information problem with a second step where we remove some locations from the ground truth datasets, which `were not observed'.

\subsubsection{Ground truth weed location model}\label{sec:GroundTruthSimModel}

The simulated ground truth weed locations
$\pppGTPattern_1,\ldots,\pppGTPattern_{\pppGTPointsNum}$ were drawn from a stochastic point-process model, where $\pppGTPointsNum$ denotes the total number of weed locations generated in the area under consideration.
More specifically,  the sequence of simulated locations
$\pppGTPattern_1,\ldots,\pppGTPattern_{\pppGTPointsNum}$ was obtained as a realization of an inhomogeneous Poisson point process
$\pppGTProcess_1,\pppGTProcess_2,\ldots$. In the following, we give a short introduction of the mathematical background and  refer to \citet{Chiu2013} for a more in-depth discussion.

Consider a bounded  sampling window $\pppSimWindow \subset \R^2$, which in our case coincides with the considered field visualized in Figure~\ref{fig:experimentalData}, and the expected number of weed locations $\Lambda(B)$  for a subset $B \subset \pppSimWindow$ (`parts of the considered field')  given by the integral  $\pppIntensityMeasure{B} = \int_B \pppIntensityFunction(x)\, \mathrm{d}x$  of a  (non-negative) intensity function $\pppIntensityFunction: \pppSimWindow \to [0, \infty)$. Then, one says that a sequence of random locations $S_1,S_2,\ldots\subset \pppSimWindow$
  follows an  inhomogeneous {\it Poisson point process}
with intensities $\Lambda(B), B\subset \pppSimWindow$, if the following  conditions are fulfilled: Consider  the random number of points $\pppCountingMeasure{B} = \#\!\left\{S_i: S_i \in B \text{ for } i = 1, 2, \dots  \right\}$ in any test set $B \subset \pppSimWindow$, where  $\#(\cdot)$ denotes set cardinality, and assume  that
\begin{enumerate}[label=(\emph{\alph*})]
    \item the random variable  $\pppCountingMeasure{B}$ is Poisson distributed, i.e.
    \[ \Pr\left( \pppCountingMeasure{B} = n\right) = \frac{{\pppIntensityMeasure{B}}^n}{n!} \exp(  -\pppIntensityMeasure{B} )\qquad\mbox{for each $n = 0, 1, \ldots$,}
    \]
    \item the random numbers of points $\pppCountingMeasure{B_1}, \dots, \pppCountingMeasure{B_k}$ in $k$ pairwise disjoint (i.e. non-overlapping) test sets $B_1, \dots, B_k \subset \pppSimWindow$ are independent of each other, for each $k=2,3,\ldots$.
\end{enumerate}
Note that condition (\emph{a}) implies that the expectation of the random number of points  $\pppCountingMeasure{B}$ in the set $B \subset \pppSimWindow$ is given by $\mathbb{E}\, \pppCountingMeasure{B}=\Lambda(B)$. Thus, indeed, $\pppIntensityMeasure{B}$ measures the expected number of weeds in a given part  $B \subset \pppSimWindow$ of the field, and the intensity function $\pppIntensityFunction: \pppSimWindow \to [0, \infty)$ governs the spatial distribution of the simulated weeds.

In Section~\ref{sec:results}, we consider various (virtual) scenarios where the weed locations have different spatial distributions. These scenarios were modeled by corresponding choices of the intensity function $\pppIntensityFunction{}$. For this, the function $\pppIntensityFunction{}$ was chosen such that the expected number of points
$\mathbb{E}\, \pppCountingMeasure{\pppSimWindow}= \int_{\pppSimWindow} \pppIntensityFunction(x) \,\mathrm{d}x$
in the sampling window $\pppSimWindow$ was set equal to the number of weed points $\pppExpPointsNum$ in the experimental ground truth dataset \datasetExperimentalData{} multiplied by some factor $\pppIntensityFactor > 0$, which we call intensity factor hereinafter, i.e., $\int_{\pppSimWindow} \pppIntensityFunction(x) \,\mathrm{d}x = \pppIntensityFactor \pppExpPointsNum$.
More precisely, we chose $\pppIntensityFunction{}$ by considering
a certain basis function $\pppBaseIntensityFunction: \pppSimWindow \to [0, \infty)$ and
a normalizing factor $\pppIntensityFunctionNormalization > 0$ such that
$\int_{\pppSimWindow}\pppIntensityFunctionNormalization \pppBaseIntensityFunction(x)\, \mathrm{d}x =  \pppExpPointsNum$. Then, in a second step, we multiply $\pppIntensityFunctionNormalization \pppBaseIntensityFunction(x)$ by the intensity factor  $\pppIntensityFactor$, i.e., $ \pppIntensityFunction(x)=\pppIntensityFactor\pppIntensityFunctionNormalization \pppBaseIntensityFunction(x)$ for each $x\in\pppSimWindow$.

\begin{figure}[ht]
    \includegraphics[width=\linewidth]{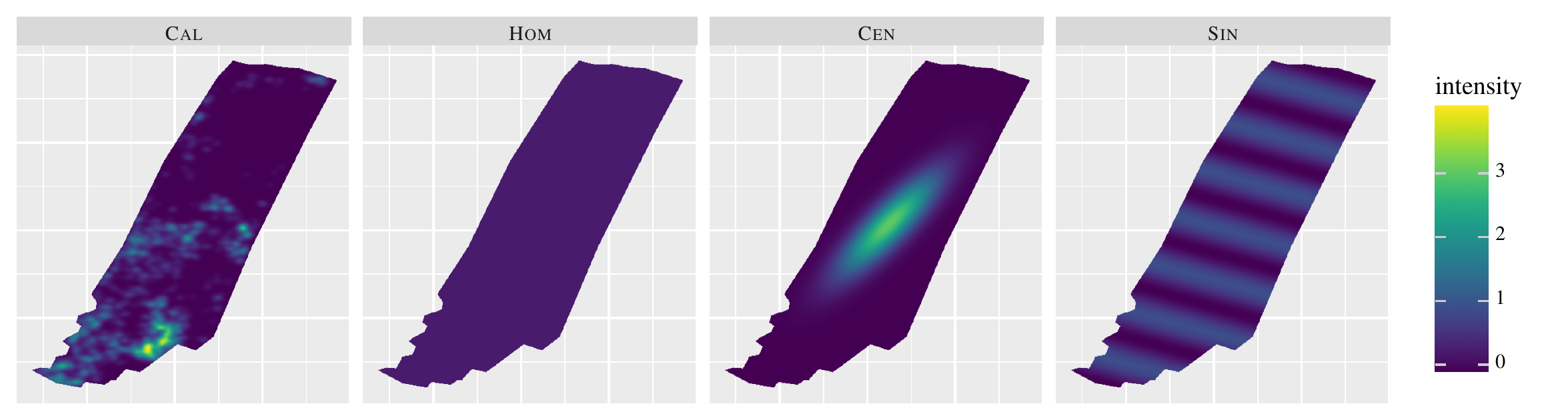}
    \caption{\label{fig:intensityFunctions}Contour plots of the intensity functions $\pppIntensityFunction$ with intensity factor $\pppIntensityFactor = 1$ of the inhomogeneous Poisson point processes used for the simulated ground truth datasets.}
\end{figure}

Note that by considering different kinds of  basis  functions
$\pppBaseIntensityFunction$ we were able to generate different types of weed distribution patterns, see Figure~\ref{fig:intensityFunctions}.  Moreover, the intensity factor $\pppIntensityFactor$  allowed us to investigate different degrees of severity of the weed infestation without changing its spatial distribution.

The stochastic ground truth dataset models---given primarily by their corresponding basis function $\pppBaseIntensityFunction$---are described in the following.

\paragraph{Calibrated ground truth model \normalfont{\datasetCalibratedInhomogeneousPoisson{}}}
The model \datasetCalibratedInhomogeneousPoisson{} was calibrated to the experimental ground truth dataset \datasetExperimentalData{} by employing a non-parametric kernel smoothing function (see, e.g., \citet{Hastie2009}) as basis functions $\pppBaseIntensityFunction$. More specifically,
        \[ \pppIntensityFunction(x) = \pppIntensityFactor \pppIntensityFunctionNormalization \sum_{i = 1}^{\pppExpPointsNum} \pppIntFunCalibratedKernel\left( \lVert x - \pppExpPattern_i \rVert \right) \quad \text{ with }  \pppIntFunCalibratedKernel(z) = \exp \left(-\frac{z^2}{2{\pppIntFunCalibratedBandwidth{}}^2} \right) \]
        for $x \in \pppSimWindow, z \geq 0$, the normalization factor $\pppIntensityFunctionNormalization$, and some bandwidth parameter $\pppIntFunCalibratedBandwidth > 0$, where $\Vert x-s\Vert$  denotes the Euclidean distance of $x,s\in\R^2$. Furthermore, $(\pppExpPattern_1, \dots, \pppExpPattern_{\pppExpPointsNum})$ are the weed locations of the experimental ground truth dataset \datasetExperimentalData{}. The bandwidth $\pppIntFunCalibratedBandwidth{}$ was chosen by drawing values at random from a gamma distribution with mean 1.5 and standard deviation 1.1. The best value was selected by considering the resulting intensity function as kernel density estimator (through normalization with $1/\pppExpPointsNum$) and maximizing the likelihood using cross-validation given the locations in \datasetExperimentalData{} \citep{Loader1999,Hastie2009}. In summary, \datasetCalibratedInhomogeneousPoisson{} thus closely models the spatial distribution of the experimental data and provides the most realistic simulated dataset, see Figure~\ref{fig:intensityFunctions}. Compared to \datasetExperimentalData{}, however, in \datasetCalibratedInhomogeneousPoisson{} the exact locations of the weeds are drawn at random and can be influenced by choosing different values for the intensity factor $\pppIntensityFactor$.

\paragraph{Homogeneous ground truth model \normalfont{\datasetSyntheticHomogeneousPoisson{}}}
The model \datasetSyntheticHomogeneousPoisson{} is based on a homogeneous Poisson process. Then, the intensity function $\pppIntensityFunction$ is constant, i.e., $\pppIntensityFunction(x) = \pppIntensityFactor \pppIntensityFunctionNormalization$ for each $x \in \pppSimWindow$ with  $\pppBaseIntensityFunction(x)=1$   for each $x \in \pppSimWindow$ and   $\pppIntensityFunctionNormalization=\pppExpPointsNum/|\pppSimWindow|$, where $|\pppSimWindow|>0$  denotes the area of the sampling window $\pppSimWindow$. This model represents the case where the weeds are located completely at random across the field $\pppSimWindow$, see Figure~\ref{fig:intensityFunctions}.

\paragraph{Centered ground truth model \normalfont{\datasetSyntheticCenteredPoisson{}}}
The weed locations of the model \datasetSyntheticCenteredPoisson{} are drawn using a bivariate normal distribution with expectation vector $\pppIntFunCenteredMean\in\pppSimWindow$ being the centroid of the sampling window $\pppSimWindow$. More precisely,
    \[
        \pppIntensityFunction(x) = \frac{\pppIntensityFactor \pppIntensityFunctionNormalization}{\sqrt{(2\pi)^2 \det \pppIntFunCenteredCov}} \exp\left(  -\frac{1}{2} (x-\pppIntFunCenteredMean)^{\operatorname{T}} (\pppIntFunCenteredCov)^{-1} (x-\pppIntFunCenteredMean) \right)
    \]
        for each $x \in \pppSimWindow$, where the positive definite dispersion matrix $\pppIntFunCenteredCov \in \R^{2 \times 2}$   was set equal to
        \[
            \pppIntFunCenteredCov = \begin{pmatrix}
                218.8 & 324.1 \\
                324.1 & 549.4
            \end{pmatrix}.
        \]
    The entries of $\pppIntFunCenteredCov$ were determined by visual examination
    and rescaled to meters. In summary, in the model \datasetSyntheticCenteredPoisson{}, the weed locations are concentrated in a central cluster around the expectation vector $\pppIntFunCenteredMean\in\pppSimWindow$, while only few weeds are generated near the boundary of the field $\pppSimWindow$, see Figure~\ref{fig:intensityFunctions}.

\paragraph{Sinusoidal ground truth model \normalfont{\datasetSyntheticSinusoidalPoisson{}}}
The intensity function $\pppIntensityFunction$ of the model \datasetSyntheticSinusoidalPoisson{} arranges the weed locations in sinusoidal waves through the sampling window, where $\pppIntensityFunction$ is given by
        \[
            \pppIntensityFunction(x) = \pppIntensityFactor \pppIntensityFunctionNormalization \operatorname{sin}\left( \frac{2\pi\left< \pppIntFunSinDirection, x \right>}{\pppIntFunSinWaveLength} \right) + 2\qquad\mbox{for each $x \in \pppSimWindow$.}
        \]
Here $<\cdot, \cdot> : \R^2 \times \R^2 \to \R$ is the dot product  with the (unit) normal vector of the wave front $\pppIntFunSinDirection \in \R^2$ and the wave length $\pppIntFunSinWaveLength > 0$.   The wave direction $\pppIntFunSinDirection$ has been set equal to the (normalized) direction of the longest side of the rectangular bounding box of the sampling window $\pppSimWindow$, and $\pppIntFunSinWaveLength = 28.3\,\meter$ was chosen by visual examination and rescaled to meters. Therefore, the dataset model \datasetSyntheticSinusoidalPoisson{} produces weed location in a wave-like pattern, see Figure~\ref{fig:intensityFunctions}.


\subsubsection{Observation model}
In practice, only some of the weeds in the population can be observed. This can have several reasons such as the plants were not yet in a growth state in which they can be detected, they were occluded by other objects, etc. For our simulation study, we imitated this by eliminating some of the weed locations from the ground truth datasets (so-called {\it independent thinning}). The remaining locations acted as the observed datasets. Here, we distinguished between two different cases analogous to the experimental dataset: the thinning probability was chosen such that the expected number of locations in the resulting dataset was (i) equal to that in \datasetExperimentalDataEarlyStandard{}, and (ii) equal to that in \datasetExperimentalDataLateStandard{}.

More formally, each of  the  weed locations
$\pppGTPattern_1,\ldots,\pppGTPattern_{\pppGTPointsNum}$ of a simulated ground truth dataset model (i.e., \datasetCalibratedInhomogeneousPoisson{}, \datasetSyntheticHomogeneousPoisson{}, \datasetSyntheticCenteredPoisson{}, or \datasetSyntheticSinusoidalPoisson{}) was eliminated with a certain probability independent of its position and the deletion of any other weed location  (so-called independent thinning, see \cite{Chiu2013}). As mentioned above, two different thinning probabilities were considered. With $\pppExpPointsNum, \pppExpEarlyPointsNum$, and $\pppExpLatePointsNum$ the number of weed locations in \datasetExperimentalData{}, \datasetExperimentalDataEarlyStandard{}, and \datasetExperimentalDataLateStandard{}, respectively, the first thinning probability is given by $\pppObservationThinningProbEarly = {\pppExpEarlyPointsNum}/{\pppExpPointsNum}$. The resulting models correspond to \datasetExperimentalDataEarlyStandard{} and are denoted with  superscript `\datasetEarlyIdentifier{}', e.g., \datasetSyntheticHomogeneousPoisson$^{\datasetEarlyIdentifier{}}$.    The second thinning probability, on the other hand, is given by $\pppObservationThinningProbLate = {\pppExpLatePointsNum}/{\pppExpPointsNum}$. The resulting models correspond to \datasetExperimentalDataLateStandard{} and are denoted with  superscript `\datasetLateIdentifier{}', e.g., \datasetSyntheticHomogeneousPoisson$^{\datasetLateIdentifier{}}$. As seen in Section~\ref{sec:ObservedExpLocations}, \datasetExperimentalDataEarlyStandard{} contained fewer observed weeds compared to \datasetExperimentalDataLateStandard{}. Naturally, this fact translates to the corresponding simulated datasets. For the formal treatment of the present simulation study both thinnings are treated analogously. For simplicity, we therefore denote the $\pppObservedPointsNum\le \pppGTPointsNum$ ``observed'' weed locations
selected from the  originally simulated weed locations
$\pppGTPattern_1,\ldots,\pppGTPattern_{\pppGTPointsNum}$  as $\pppObservedPattern_1, \dots, \pppObservedPattern_{\pppObservedPointsNum}$.

\subsection{Treatment strategies}
Now that the generation of the simulated datasets is established, we describe two different treatment strategies, which are applied in Section~\ref{sec:results} to the experimental and simulated datasets. The treatment strategies are a combination of three parts: (i) the action threshold (see Section~\ref{sec:action_threshold}) to select the targeted weed locations, (ii) the treatment tool, which is either an autonomous robot (see Section~\ref{sec:robot}), or a tractor with attachment (see Section~\ref{sec:tractor}), and (iii) the (parameter) configuration of the virtual treatment tool. The considered tools are based on the real-world analogues presented in \citet{Stoll2020} and \citet{Martin2022}.

\subsubsection{Action threshold}\label{sec:action_threshold}
For both  treatment strategies considered in this paper, we apply an \emph{action threshold}, which is used as a simple preprocessing method to improve the treatment instead of simply targeting each observed weed locations. More specifically, we ignore isolated weeds, which might drastically increase the distance a treatment tool has to drive, while having only a very limited effect on the treatment quality (see Section~\ref{sec:resultsGeneralSetting} for a description of  suitable performance measures). Furthermore, in Section~\ref{sec:results} varying settings for the action threshold are investigated.

Formally, the action threshold $\strategyParamActionThreshold > 0\,\meter$  caps the nearest neighbor distance (in meter) of the treated weed locations, i.e., given the observed weed locations $\pppObservedPattern_1, \dots, \pppObservedPattern_{\pppObservedPointsNum}$, the $\pppTargetedPointsNum\le\pppObservedPointsNum$ targeted  weed locations $\pppTargetedPattern_1, \dots, \pppTargetedPattern_{\pppTargetedPointsNum}$ that are fed into the treatment tools are those points $\pppObservedPattern_i$ with $i = 1, \dots, \pppObservedPointsNum$ such that there is some $j \neq i$ with $\lVert \pppObservedPattern_i - \pppObservedPattern_j \rVert \leq \strategyParamActionThreshold$. Note that by setting $\strategyParamActionThreshold = \infty\,\meter$, all observed weed locations are being targeted.

\begin{figure}[ht]
    \begin{subfigure}[t]{0.47\linewidth}
        \includegraphics[width=\linewidth]{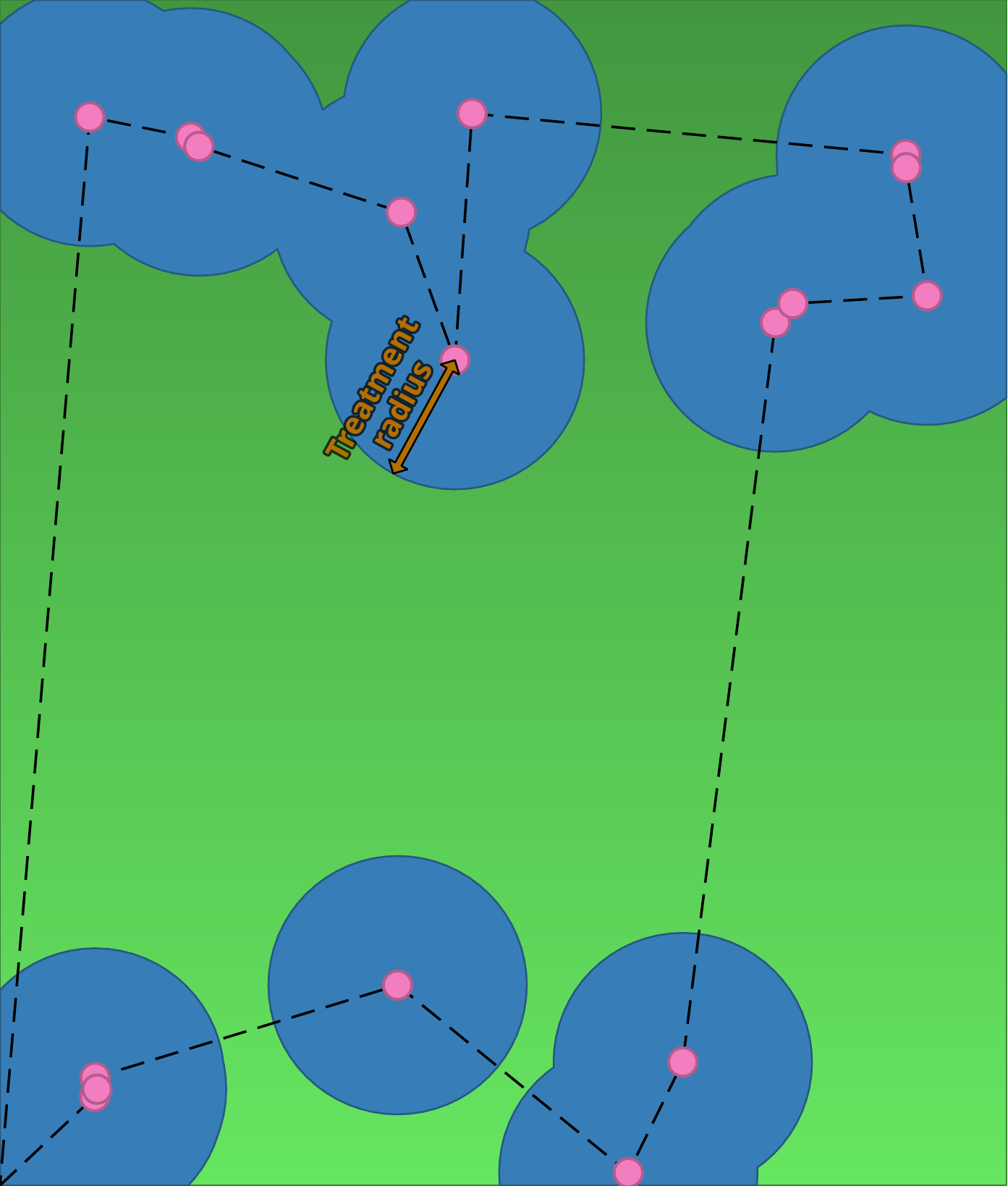}
        \caption{\label{fig:strategyRobot}The tool \strategyLoneRobot{}  goes directly from one targeted weed location to the next, where a disk-shaped area is treated.}
    \end{subfigure}\hfill
    \begin{subfigure}[t]{0.47\linewidth}
        \includegraphics[width=\linewidth]{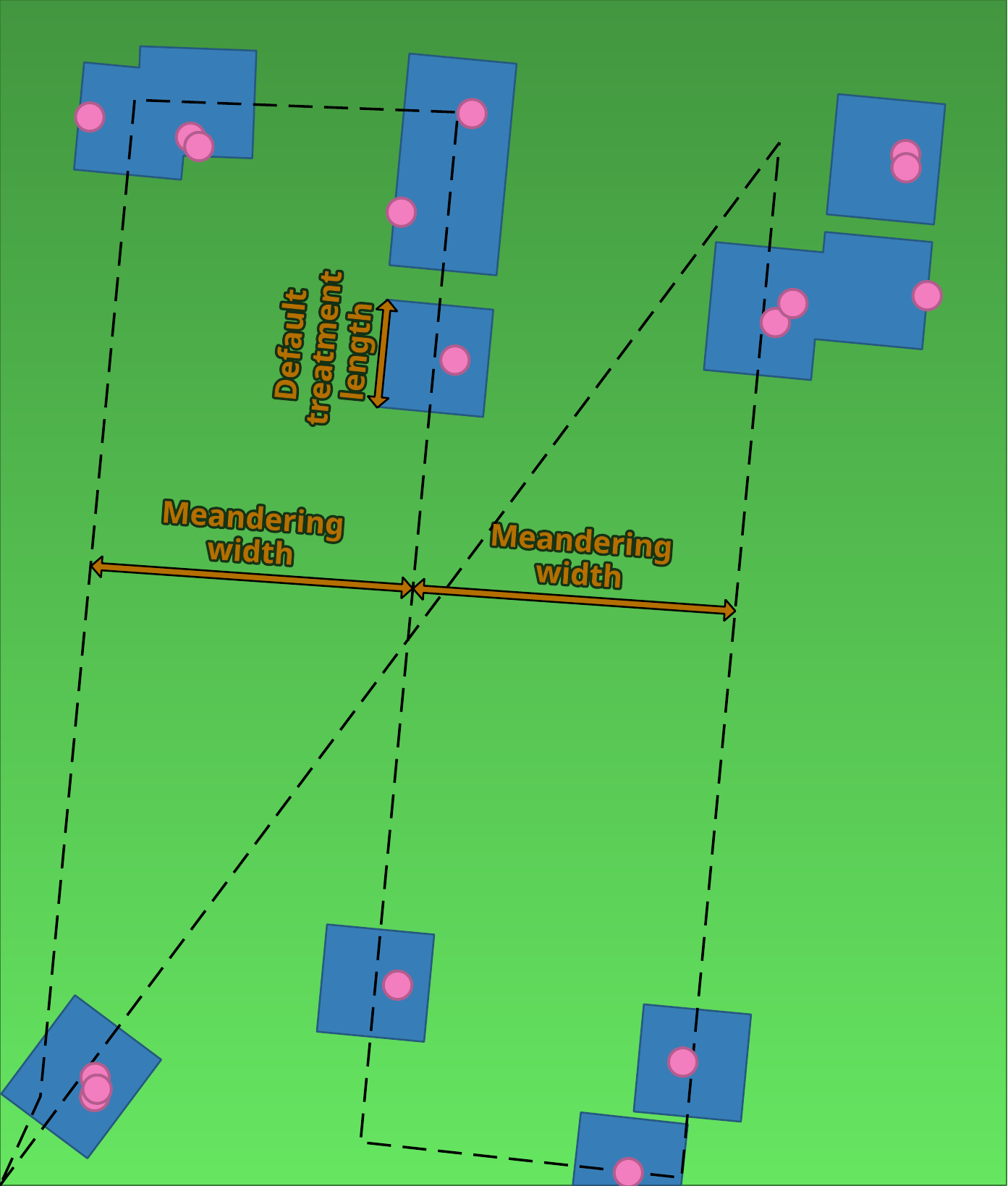}
        \caption{\label{fig:strategyTractor}The tool \strategySnakingTractor{} meanders through the weed infested area such that each targeted weed location is covered by one of its separately controllable treatment sections. The meandering width is thus equal to the width of the attachment. Once the tool crosses a targeted weed location a rectangular area with a given side length along the driving direction is treated. In the illustrated example the treatment tool consists of three sections.}
    \end{subfigure}
    \caption{Illustration of the  treatment tools \strategyLoneRobot{} (left) and \strategySnakingTractor{} (right) and their parameters. The treatment tools start at a starting point $\strategyStartingPoint$ (bottom left corner) and drive over the field $\pppSimWindow$ (green area) along a specified route (dashed line). Every time it crosses a targeted weed location (pink dots), a tool-specific area is treated (blue area).}
\end{figure}

\subsubsection{Autonomous robot}\label{sec:robot}
The first treatment tool, denoted as \strategyLoneRobot{}, imitates an autonomous robot that is able to drive directly from one target location to another (see, e.g., \citet{Stoll2020}). The treatment is performed, e.g., by activating a circular mower. In reality, this would result in an oblong treatment area along the driving direction as no cutting occurs in the center of the mower. For simplicity sake, however, we assume a disk-shaped area. After the treatment, the robot continues straight to the next target location, see Figure~\ref{fig:strategyRobot}.

The parameter configuration consists of a treatment radius $\strategyParamTreatmentRadius > 0\,\meter$ (in meter), which specifies the radius of the treated circular area centered at each targeted weed location $\pppTargetedPattern_1, \dots, \pppTargetedPattern_{\pppTargetedPointsNum}$. For our simulation study we assume that both treatment tools  start in a fixed point $\strategyStartingPoint \in \pppSimWindow$, which will also be the point to where they return after they have finished. The route of the treatment tool \strategyLoneRobot{} through the field is determined as the shortest tour through all target locations $\pppTargetedPattern_1, \dots, \pppTargetedPattern_{\pppTargetedPointsNum}$ and the starting point $\strategyStartingPoint$. This is known as the (Euclidean) traveling salesman problem \citep{Jungnickel2008}, which we solved using the \emph{OR-Tools} library \citep{ortools}. The library produced a not necessarily optimal, but reasonably good route as finding the optimum can be very time consuming. The initial tour was chosen by iteratively adding the weed location with the shortest distance to the previous location beginning with the starting location. It is worth mentioning that as a consequence the route is independent from the treatment radius $\strategyParamTreatmentRadius$. More sophisticated strategies would also have been possible where instead of targeting observed weed locations directly, artificial target locations could have been computed depending on $\strategyParamTreatmentRadius$ such that multiple (observed) weeds were treated simultaneously. However, these lead to extra complexity through additional constraints in practice such as through the imperfect positioning accuracy on the field.

\subsubsection{Tractor with attached treatment tool}\label{sec:tractor}
The second treatment tool, denoted as \strategySnakingTractor{}, behaves like a tractor with an attached treatment tool (such as the water-hydraulic tool proposed in \citet{Stoll2020} and \citet{Martin2022}) that covers the weed infested area in a winding path. The attachment consists of several separately engageable sections for a site-specific treatment. Each of these sections treat a rectangular area surrounding a targeted weed location, see Figure~\ref{fig:strategyTractor}.

For the precise definition of \strategySnakingTractor{}, we consider the targeted weed locations $\pppTargetedPattern_1, \dots, \pppTargetedPattern_{\pppTargetedPointsNum}$. The weed infested area $\strategyInvestedArea \subset \pppSimWindow$ is then given by the convex hull \citep{Berg2008} of these targeted weed locations, and the primary driving direction $\strategyTractorPrimaryDrivingDirection \in \R^3$ with $\lVert \strategyTractorPrimaryDrivingDirection \rVert = 1$ is the direction of the longest side of the (arbitrarily oriented) minimum rectangular bounding box of $\strategyInvestedArea$ (or equivalently of the targeted weed locations). As described in Section~\ref{sec:robot} for \strategyLoneRobot{}, the tractor  starts in a fixed point $\strategyStartingPoint$. From there it traverses  $\strategyInvestedArea$ primarily in parallel line segments determined by the vector $\strategyTractorPrimaryDrivingDirection$. The distance between these line segments is given by the meandering width $\strategyParamMeanderingWidthInM > 0\,\meter$ (in meter), which is equivalent to the width of the attached treatment tool. Relative to $\strategyStartingPoint$, the farthest parallel line segment is located such that its distance to the farthest targeted weed location (which is on the boundary of $\strategyInvestedArea$) is equal to $\strategyParamMeanderingWidthInM / 2$. The tractor crosses orthogonally from one line segment to the other such that the infested area $\strategyInvestedArea$ is fully covered (i.e., the distance to the boundary of $\strategyInvestedArea$ is at most $\strategyParamMeanderingWidthInM/2$). The positive turning radius of a real tractor is neglected in this model. After $\strategyInvestedArea$ has been traversed, the tractor returns to the starting point $\strategyStartingPoint$. Note that there are two possibilities in which direction to start the traversal, namely $\strategyTractorPrimaryDrivingDirection$ or $-\strategyTractorPrimaryDrivingDirection$. We chose the one that leads to the shortest route. An example route for \strategySnakingTractor{} can be seen in Figure~\ref{fig:strategyTractor}.

While the tractor moves along the route described above, the $\strategyParamToolSegmentCount > 0$ individual sections of the attached treatment tool perform a site-specific weed control.
For this, the line segment of length $\strategyParamMeanderingWidthInM$ perpendicular to the driving route is divided into $\strategyParamToolSegmentCount$ parts. These parts correspond to the sections of the treatment tool, which engage (at least) $\strategyParamDefaultTreatmentLength / 2$ in front of and disengage (at least) $\strategyParamDefaultTreatmentLength / 2$ behind a targeted weed location with the default treatment length $\strategyParamDefaultTreatmentLength > 0\,\meter$. This leads to rectangular treatment areas for isolated weeds with side lengths $\strategyParamDefaultTreatmentLength \times \strategyParamMeanderingWidthInM / \strategyParamToolSegmentCount$. Note, however, that for targeted weed locations close to each other these rectangles can merge. For the simulation study considered in Section~\ref{sec:results}, we set the meandering width equal to $\strategyParamMeanderingWidthInM = 2.5\,\meter{}$ and the default treatment length to $\strategyParamDefaultTreatmentLength = \strategyParamMeanderingWidthInM / \strategyParamToolSegmentCount$. The number of sections $\strategyParamToolSegmentCount$ will be varied.

\section{Results}\label{sec:results}
Now that the entire simulation framework, i.e. the  generation of the simulated weed locations and the treatment strategies,
has been laid out, we describe the results of our case study.

\subsection{General setting}\label{sec:resultsGeneralSetting}
The general setup for the case study is as follows. First, a set of ground truth weed locations $\{ \pppGTPattern_1, \dots, \pppGTPattern_{\pppGTPointsNum} \}$  and the observed subset $\{ \pppObservedPattern_1, \dots, \pppObservedPattern_{\pppObservedPointsNum} \}$ with $\pppGTPointsNum \geq \pppObservedPointsNum \geq 0$ were obtained, either from the experimental datasets (see Sections~\ref{sec:GroundTruthExpLocations} and \ref{sec:ObservedExpLocations}, respectively) or as realizations from stochastic models (see Section~\ref{sec:simulatedData}). Recall from Section~\ref{sec:GroundTruthSimModel} that for the simulated datasets, it is possible to vary the mean number of ground truth weed locations relative to the experimentally observed ones in \datasetExperimentalData{} through the intensity factor $\pppIntensityFactor$. In the following, we consider three different cases, i.e. $\pppIntensityFactor \in \{ 0.5, 1, 2 \}$, which are denoted with a subscript `$\datasetHalfIdentifier$' (such as \datasetCalibratedInhomogeneousPoissonEarlyHalf{}), `$\datasetStandardIdentifier$' (such as \datasetCalibratedInhomogeneousPoissonEarlyStandard{}), or `$\datasetDoubleIdentifier$' (such as \datasetCalibratedInhomogeneousPoissonEarlyDouble{}), respectively. Moreover, two observations were simulated, where either fewer or more weed locations were observed, just like with the experimental datasets \datasetExperimentalDataEarlyStandard{} and \datasetExperimentalDataLateStandard{}, respectively. Based on the observed weed locations, the $\pppTargetedPointsNum \geq 0$ targeted weed locations $\pppTargetedPattern_1, \dots, \pppTargetedPattern_{\pppTargetedPointsNum}$ were selected by applying an action threshold $\strategyParamActionThreshold \in \{ 2.5\,\meter, 5\,\meter, \infty\,\meter \}$ eliminating locations with a nearest neighbor distance larger than $\strategyParamActionThreshold$. Note that for $\strategyParamActionThreshold = \infty\,\meter$ no locations were thus removed from the observed dataset. In total six different configurations of treatment tools were considered, three for \strategyLoneRobot{} with treatment radius $\strategyParamTreatmentRadius \in \{  0.2\,\meter, 0.4\,\meter, 1.25\,\meter \}$ and three for \strategySnakingTractor{} with $\strategyParamToolSegmentCount \in \{ 1, 5, 10 \}$ separately controllable sections. Only the targeted weed locations were given to each of these individual strategies, which then produced a set of treated weed locations $\{ \resTreatedPattern_1, \dots, \resTreatedPattern_{\resTreatedPointsNum} \}$ with $\pppGTPointsNum \geq \resTreatedPointsNum \geq \pppTargetedPointsNum$ and a treated subset $\resTreatedArea$ of the field  $\pppSimWindow$. From the definition of the treatment tools, it is clear that every targeted weed location was treated, but there might have been some weed locations near a targeted location that were also treated as `collateral damage'. The point where the treatment tools started and finished their tour $\strategyStartingPoint$ was set to the lowest point on the left of the field $\pppSimWindow$ and remained fixed for all simulations.


\subsection{Performance measures}

In order to quantify the performance of a treatment, we computed the following performance measures, which are minimized by an opimal treatment:
\begin{enumerate}[label=(\emph{\alph*})]
    \item the distance \metricDrivingDistanceInM{} driven by the treatment tool  (in meter, including the distance from and to the starting point $\strategyStartingPoint$),
    \item the number \metricFracRemainingWeeds{}  of remaining weed locations relative to the total number of ground truth plants, where $\metricFracRemainingWeeds{} = \frac{\pppGTPointsNum - \resTreatedPointsNum}{\pppGTPointsNum}$,
    \item  the maximum density  $\metricHighestRemainingWeedsDensityM{2}$ of remaining weeds in a disk of radius 2\,\meter, where
        \[ \metricHighestRemainingWeedsDensityM{2} = \frac{1}{4\,\squareMeter\ \pi} \  \operatorname*{max}_{x \in \metricHighestDensityGrid} \#\left\{ i: \lVert \pppGTPattern_i - x \rVert \leq 2\,\meter \text{ and } \pppGTPattern_i \not\in \{ \resTreatedPattern_1, \dots, \resTreatedPattern_{\resTreatedPointsNum} \} \right\} \]
        and $\metricHighestDensityGrid \subset \pppSimWindow$ is a square 5\,\centimeter-grid of the field $\pppSimWindow$,
    \item the treated area $\metricTreatedAreaFrac$ relative to the area of the whole field, where $\metricTreatedAreaFrac = \frac{|\resTreatedArea|}{|\pppSimWindow|}$ and $|\cdot|$ denotes the area of a given set, and
    \item the treated area $\metricTreatedAreaPerWeed{}$ per treated weed (treatment efficiency), where $\metricTreatedAreaPerWeed{} = \frac{|\resTreatedArea|}{\resTreatedPointsNum}$ in \squareMeter.
\end{enumerate}
Note that the radius of the disk for $\metricHighestRemainingWeedsDensityM{2}$ was chosen as a trade-off between capturing varying local weed concentrations, while still being large enough to cover more than one weed in a cluster.

A great advantage of simulated data drawn from stochastic weed location models, compared to experimental data, is that the virtual treatments can be repeated without changing the considered scenario (in terms of the expected number of weed locations, their spatial distribution, etc.). Through these replications independent samples of the treatment results can be obtained, which leads to great statistical reliability. For this  reason, we drew 10 samples from each of the stochastic simulation models. In the following only the mean values obtained for the quality measures ({\it a}) -- ({\it e}) are presented.

\subsection{Comparison of treatment strategies}

The first question, we want to answer is which strategy is the best one. Unfortunately, since no strategy outperforms its alternatives with respect to all performance measures  ({\it a}) -- ({\it e})
in all scenarios, the definition of optimality has to be relaxed. The importance of a performance measure might vary from context to context. We therefore aim to investigate common trade-offs, which might support decision making by identifying strictly inferior strategies.

This problem can be tackled using the so-called {\it Pareto-optimality} known from multiobjective optimization, see, e.g., \citet{Miettinen2012}. In our case a treatment strategy is Pareto-optimal if every other strategy with a better outcome regarding one performance measure would have a worse result regarding another measure. If for a strategy, on the other hand, there is no performance measure where it surpasses the others, this strategy is clearly inferior as the outcome can be improved regardless of what trade-off a decision maker is willing to make. This means more formally that when minimizing the objective functions (in our case performance measures) $f_1, \dots, f_k: \R^d \supset D \to \R$ for some integers $d,k$, the decision vector (in our case strategy) $x^* \in D$ is Pareto-optimal if there is no other vector $x \in D$ with $f_i(x) \leq f_i(x^*)$ for all $i = 1, \dots, k$ and $f_j(x) < f_j(x^*)$ for at least one index $j$ \citep{Miettinen2012}.

In the following, we consider only two performance measures at a time as more would result in practically all strategies being Pareto-optimal. Moreover, visualizing three or more measures  simultaneously is much harder. The results of each treatment strategy with respect to the considered performance measures are shown in a scatter plot. The criterion for the Pareto-optimality is illustrated as a line---the so-called Pareto frontier---that separates the Pareto-optimal strategies on the line from the inferior ones on the top right of the line.

\begin{figure}
    \centering
    \includegraphics[width=\textwidth]{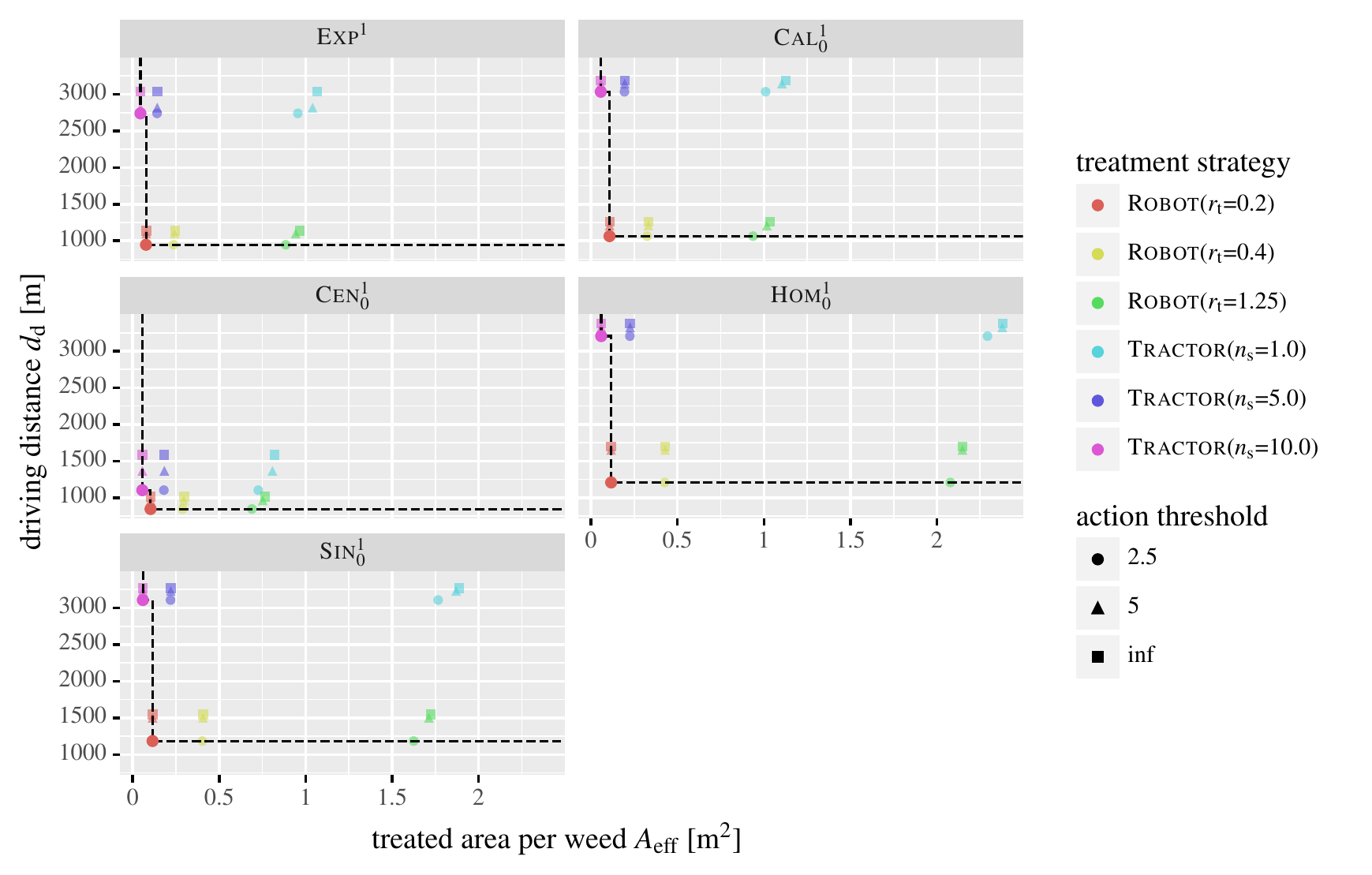}
    \caption{Scatter plots together with the corresponding Pareto frontiers of the driving distance \metricDrivingDistanceInM{} and the treated area per treated weed \metricTreatedAreaPerWeed{} for various treatment strategies and infestation scenarios.}
    \label{fig:pareto_area_per_weed_driving_dist}
\end{figure}

In order to reduce complexity, we only consider the datasets corresponding to the early observation date (of September 3rd, 2019) and set the intensity factor $\pppIntensityFactor = 1$ for the simulated data  (i.e. \datasetExperimentalDataEarlyStandard, \datasetCalibratedInhomogeneousPoissonEarlyStandard, \datasetSyntheticHomogeneousPoissonEarlyStandard, \dots) when investigating the Pareto-optimality of the strategies. In Figure~\ref{fig:pareto_area_per_weed_driving_dist} the driving distance $\metricDrivingDistanceInM$ versus the treated area per treated weed  $\metricTreatedAreaPerWeed$ are shown. Here it turned out that the Pareto-optimal strategies were those with the smallest individual treatment area (i.e., the smallest considered treatment radius $\strategyParamTreatmentRadius{} = 0.2\,\meter$ for \strategyLoneRobot{}, or the maximum number of sections $\strategyParamToolSegmentCount = 10$ for \strategySnakingTractor{}). Furthermore, compared to \strategyLoneRobot{}, \strategySnakingTractor{} drove a much larger distance, but treated a slightly smaller area per weed. The scenario \datasetSyntheticCenteredPoissonEarlyStandard{} was the only one, however, where there was almost no difference between the two tools.  Note that the driving distance \metricDrivingDistanceInM{} does not change when varying the parameters \strategyParamTreatmentRadius{} or \strategyParamToolSegmentCount{} as the route of a treatment tool depends only on its type (\strategyLoneRobot{} or \strategySnakingTractor{}) and the targeted weed locations, which are a function of the action threshold \strategyParamActionThreshold{}. Regarding \strategyParamActionThreshold{}, lower values, which reduced the number of targeted weed locations, led to lower driving distances as might have been expected. Consequently, $\strategyParamActionThreshold{} = 2.5\,\meter$ produced the lowest values of \metricDrivingDistanceInM{}.

\begin{figure}
    \centering
    \includegraphics[width=\textwidth]{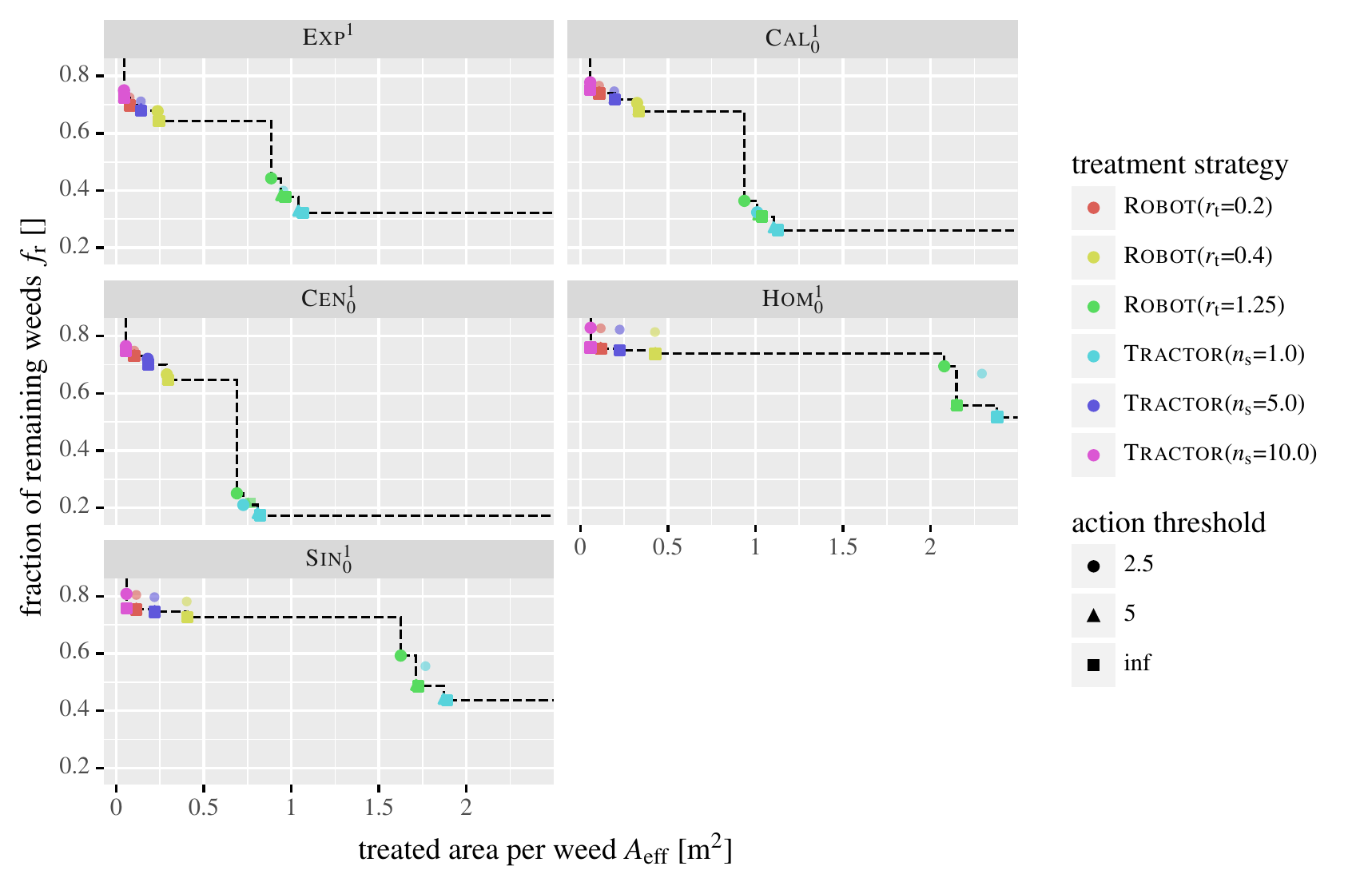}
    \caption{Scatter plots together with the corresponding Pareto frontiers of the fraction of remaining weeds \metricFracRemainingWeeds{} and the treated area per treated weed \metricTreatedAreaPerWeed{} for various treatment strategies and infestation scenarios.}
    \label{fig:pareto_area_per_weed_remaining_weeds}
\end{figure}

In Figure~\ref{fig:pareto_area_per_weed_remaining_weeds}, the trade-off between the fraction of remaining weeds \metricFracRemainingWeeds{} and the treated area per treated weed \metricTreatedAreaPerWeed{} are illustrated. Here, apparently almost all strategies were Pareto-optimal which points to the strong antagonistic dependency between the treated area and the number of remaining weeds. Two groups of treatment strategies could be made out: the first one comprised the strategies with the largest individual treatment area (i.e.  \strategyLoneRobot{} with $\strategyParamTreatmentRadius{} = 1.25\,\meter$ and \strategySnakingTractor{} with $\strategyParamToolSegmentCount = 1$) and treated significantly more area per weed at the benefit of hitting a larger percentage of the whole weed population compared to the remaining strategies in the second group. Interestingly enough, no big difference between the two tools was visible, only between their configuration. The differences between $\strategyParamActionThreshold = \infty\,\meter$ and $\strategyParamActionThreshold = 5\,\meter$ were negligible but there was visible contrast compared to $\strategyParamActionThreshold = 2.5\,\meter$. As opposed to Figure~\ref{fig:pareto_area_per_weed_driving_dist}, the latter action threshold usually performed the worst.

\begin{figure}
    \centering
    \includegraphics[width=\textwidth]{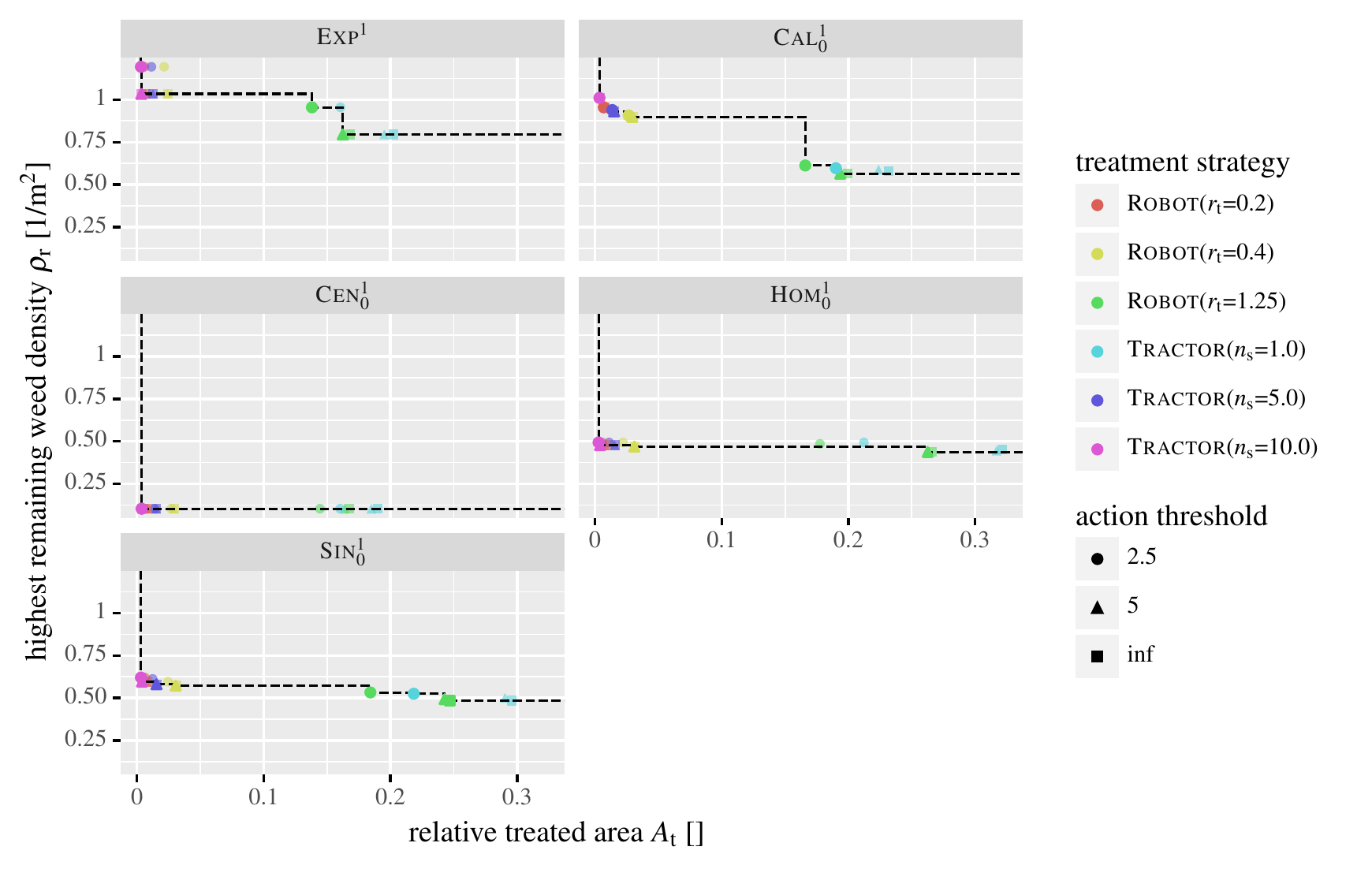}
    \caption{Scatter plots together with the corresponding Pareto frontiers of the relative treated area \metricTreatedAreaFrac{} and the highest remaining weed density \metricHighestRemainingWeedsDensityM{2} for various treatment strategies and infestation scenarios.}
    \label{fig:pareto_treated_area_weed_density}
\end{figure}

Another important pair of competing performance measures is the relative treated area \metricTreatedAreaFrac{} and the highest remaining weed density \metricHighestRemainingWeedsDensityM{2}. The corresponding results are shown in Figure~\ref{fig:pareto_treated_area_weed_density}. Here, \strategySnakingTractor{} with $\strategyParamToolSegmentCount = 10$ was Pareto-optimal in all scenarios. The other strategies, \strategyLoneRobot{} with $\strategyParamTreatmentRadius = 1.25\,\meter$ most of all, were in some cases able to decrease the highest remaining weed density \metricHighestRemainingWeedsDensityM{2}. The biggest difference was achieved for \datasetExperimentalDataEarlyStandard{} and \datasetCalibratedInhomogeneousPoissonEarlyStandard{}, whereas for the remaining scenarios little or no improvement was obtained with respect to \metricHighestRemainingWeedsDensityM{2}. Generally speaking a smaller value for the action threshold \strategyParamActionThreshold{} yielded better results.

\begin{figure}
    \centering
    \includegraphics[width=\textwidth]{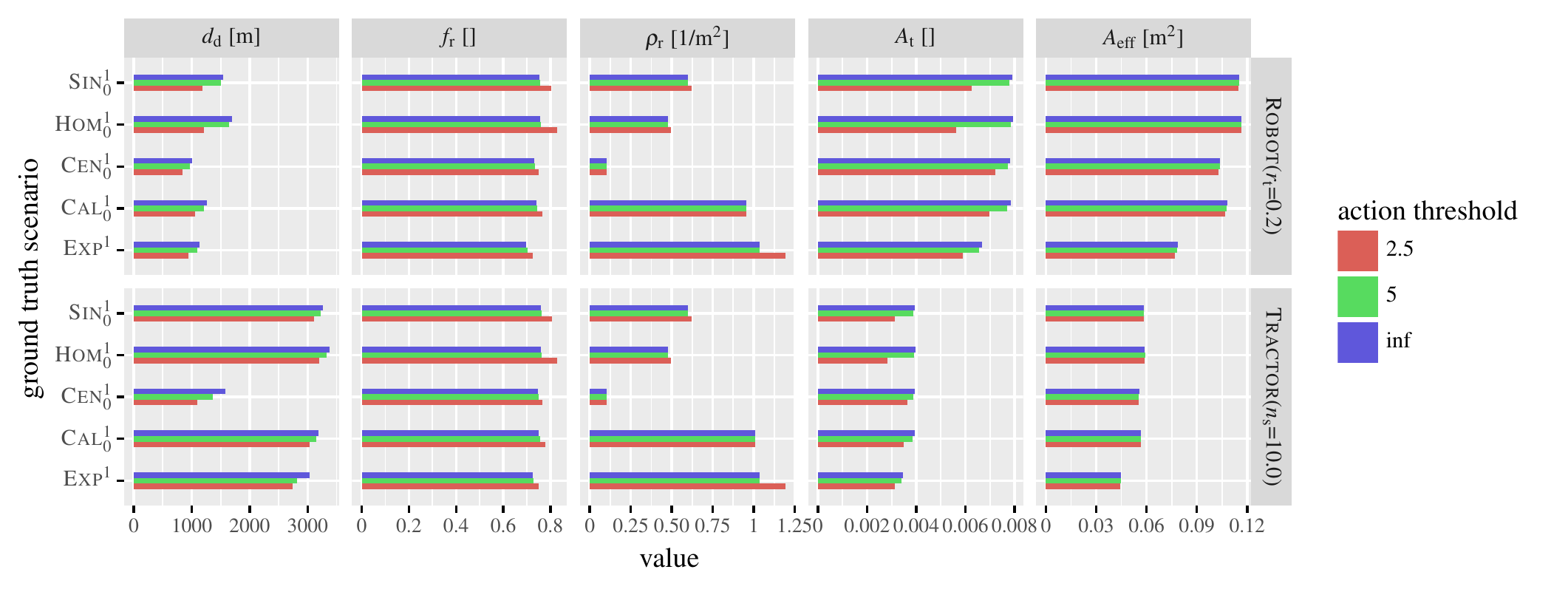}
    \caption{Influence of the action threshold $\strategyParamActionThreshold$ on the considered performance measures for various infestation scenarios.}
    \label{fig:action_threshold}
\end{figure}

\subsection{Influence of action threshold on treatment performance}

Another question that we want to investigate is how the action threshold \strategyParamActionThreshold{} affected the treatment performance. Like above, we considered only the early observation date and set the intensity factor $\pppIntensityFactor = 1$, but restricted the tool configurations to the ones described in \cite{Stoll2020} and \citet{Martin2022}, namely $\strategyLoneFarmer{}$ with treatment radius $\strategyParamTreatmentRadius = 0.2\,\meter$ and $\strategySnakingTractor$ with section count $\strategyParamToolSegmentCount = 10$. The results are shown in Figure~\ref{fig:action_threshold}.  By setting the action threshold $\strategyParamActionThreshold = 5\,\meter$, small reductions in the driving distance \metricDrivingDistanceInM{} compared to the baseline $\strategyParamActionThreshold = \infty\,\meter$ could be observed, but practically no change for the other metrics. So a small net win could be achieved. For the smallest action threshold, $\strategyParamActionThreshold = 2.5\,\meter$, more weeds remained untreated. It is also noteworthy, that especially for the \strategySnakingTractor{} the treated area per weed \metricTreatedAreaPerWeed{} was mostly unaffected by the action threshold.

\begin{figure}
    \centering
    \includegraphics[width=\textwidth]{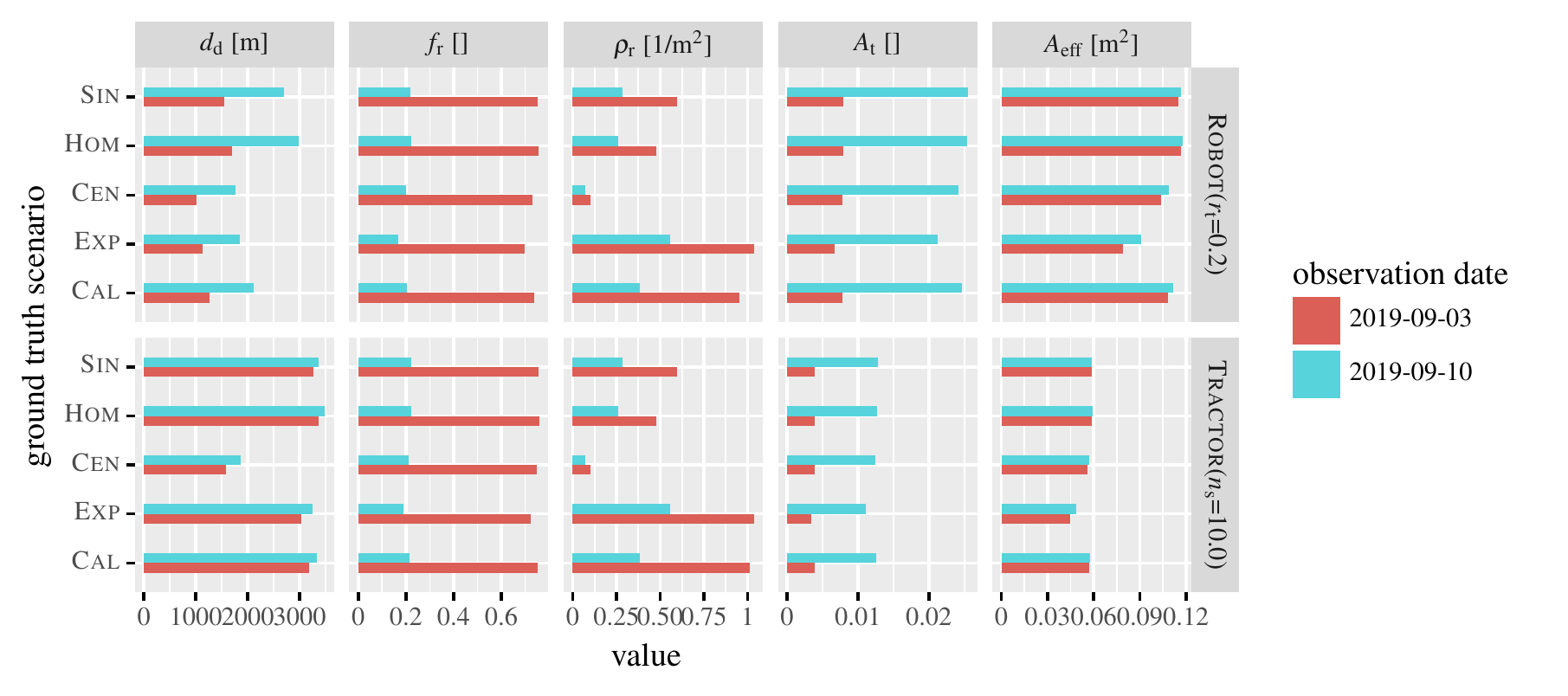}
    \caption{Influence of the observation date on the considered performance measures for various infestation scenarios.}
    \label{fig:observation_date}
\end{figure}

\subsection{Effect of observation date}
In order to study the effect of the observation date, we set the intensity factor $\pppIntensityFactor = 1$ and considered only the action threshold $\strategyParamActionThreshold = \infty\,\meter$ for the two treatment configurations, $\strategyLoneFarmer{}$ with treatment radius $\strategyParamTreatmentRadius = 0.2\,\meter$ and $\strategySnakingTractor$ with section count $\strategyParamToolSegmentCount = 10$. The resulting metrics are visualized in Figure~\ref{fig:observation_date}. Between the two observation dates, large differences in almost all considered metrics could be observed. The datasets corresponding to September 10th, 2019, where more weeds were observed ($\pppExpLatePointsNum = 1792$ versus $\pppExpEarlyPointsNum = 550$ for September 3rd, 2019) had a much better treatment result. Notable outliers were, however, the driving distance \metricDrivingDistanceInM{} for \strategySnakingTractor{}, and the treated area per treated weed for both tool, which changed only marginally. It should be kept in mind that the individual treatment areas differ and except for the driving distance \metricDrivingDistanceInM{} a comparison between the \strategySnakingTractor{} and the \strategyLoneRobot{} would not be reasonable. All ground truth scenarios produced the same qualitative behavior of the considered performance measures.

\begin{figure}[ht]
    \centering
    \includegraphics[width=\textwidth]{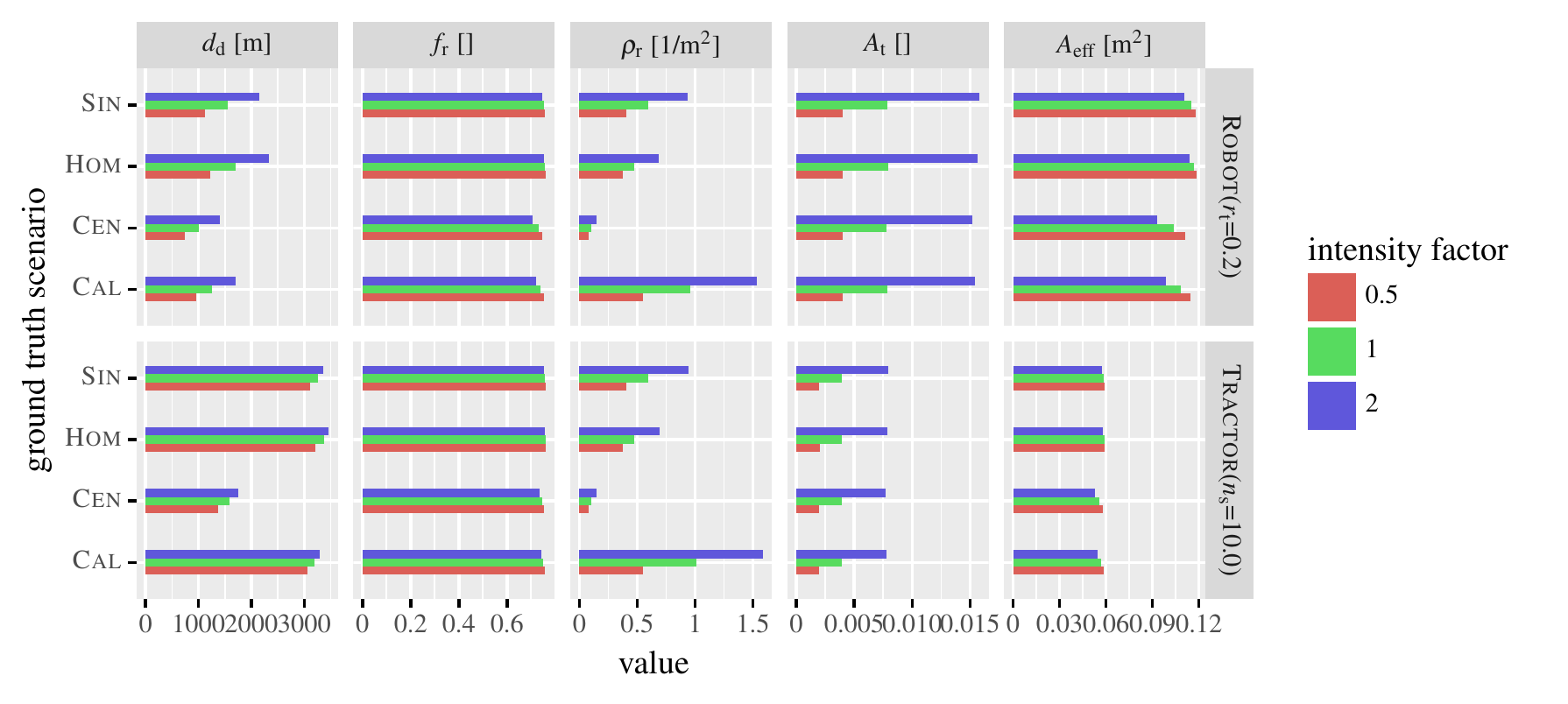}
    \caption{Influence of the intensity factor $\pppIntensityFactor$ on the considered performance measures for various infestation scenarios.}
    \label{fig:intensity_factor}
\end{figure}

\subsection{Influence of the intensity factor\texorpdfstring{ $\pppIntensityFactor$}{}}

In Figure~\ref{fig:intensity_factor} the dependency of the intensity factor $\pppIntensityFactor$ on the considered performance measures is shown, which illuminates the question how the treatment strategies performed when the number of (ground truth) weed locations varies.
Of course, only simulated datasets could be considered. As before, we focused on $\strategyLoneFarmer{}$ with treatment radius $\strategyParamTreatmentRadius = 0.2\,\meter$ and $\strategySnakingTractor$ with section count $\strategyParamToolSegmentCount = 10$, and we ignored the action threshold, i.e. $\strategyParamActionThreshold = \infty\,\meter$. Apparently, the driving distance \metricDrivingDistanceInM{} did not scale linearly with the mean number of weeds on the field. So, a low number of weeds lead to much larger distances (and therefore costs) per weed compared to cases where this number was already quite high. The \strategySnakingTractor{} tool was less affected by the increase in \metricDrivingDistanceInM{} than \strategyLoneRobot{}. For the treated area per treated weed \metricTreatedAreaPerWeed{}, it was also observable that
\metricTreatedAreaPerWeed{} decreased with increased intensity factor $\pppIntensityFactor$. So the efficiency rose as more and more unobserved weeds stood near targeted weed locations and were treated.

\section{Discussion}

When it comes to evaluating the different treatment strategies, the Figures~\ref{fig:pareto_area_per_weed_driving_dist}, \ref{fig:pareto_area_per_weed_remaining_weeds}, and \ref{fig:pareto_treated_area_weed_density} showed how difficult it is to give general suggestions. However, certain strategies can be eliminated when
considering only a few performance measures that are the most important ones in the current context. Especially when keeping the results shown in Figure~\ref{fig:intensity_factor} in mind, the \strategyLoneRobot{} strategy generally performed quite well. Only with an increasing number of weeds  \strategySnakingTractor{} obtained better results in particular if the weeds are clustered. For real world applications, additional constraints come into play, such as a limited fuel tank of an automated robot resulting in a shorter reach, or the varying costs of personal. The presented approach could also be extended to serve as a framework to investigate such more intricate questions that arise from planning a treatment tool to managing weeds on a large number of fields with a limited number of available treatment tools.

It turns out that by employing a (finite) action threshold, the driving distance \metricDrivingDistanceInM{} could be reduced. For the smallest value $\strategyParamActionThreshold{} = 2.5\,\meter$, however, a noticeable degradation of the treatment performance could be observed,
see Figure~\ref{fig:action_threshold}. The value $\strategyParamActionThreshold{} = 5\,\meter$ might be used as a compromise between a smaller driving distance and practically the same treatment performance compared to omitting the action threshold. For future research it might be worthwhile to implement more sophisticated methods to determine the set of targeted weed locations. For example, using point processes would allow for a model based prediction of unobserved weed locations. With this, additional virtual locations could be created in areas where unobserved weeds are assumed.  Another approach would be to predict unobserved weed locations using machine learning techniques. However, both approaches would require additional experimental data to build accurate models.

In Figure~\ref{fig:observation_date}, a strong dependency of the treatment success on the number of observed weeds can be seen. The importance of this relationship is only increased, when keeping in mind that only for simplicity's sake we assumed the experimental locations \datasetExperimentalData{} comprised the whole weed population. In reality, however, there might have been much more weeds on the field that were not captured during the two drone mappings. This partial information problem affects both online and offline treatment methods that either identify the targeted weed locations while the treatment tool is traversing the field (e.g., through attached cameras and real-time image processing), or decouple the acquisition of the image data/detection of weed locations with their treatment, respectively. For offline methods the weed population might change between the observation and the treatment (e.g., drone imaging in fall where the weeds are easiest to identify versus treatment in spring where the plants are most vulnerable). This might lead to an even larger amount of discrepancy between the observed and the ground truth weed locations. However, an in-depth analysis of this phenomenon is beyond the scope of the present paper. For online methods, on the other hand, it is not possible to combine information from several observations, e.g., from different days. These additional data sources could be necessary to obtain an accurate survey of the whole weed population. Hybrid approaches that combine the benefits of online and offline techniques could be a solution.

\section{Conclusions}
In the present paper, a simulation framework for investigating questions about weed control strategies was developed. A combination of experimental and simulated datasets has been employed to analyze the trade-offs between treatment strategies. Additionally, the effect of partial observation of the weed population and of different severities of the infestations on the treatment performance has been studied.

Various extensions to the proposed framework are possible. First off, more sophisticated point-process models like (Poissonian) cluster point processes for the ground truth weed location and dependent thinning for the observed subset could be investigated to obtain more variability and more complex interdependencies in the resulting point patterns. Ideally, a thorough stochastic model of real weed locations could be built. However, this would require more experimental data from different fields at several time points. Another point for enhancement would be the addition of further treatment tools and configurations.

\section*{Acknowledgements}
The project is supported by funds of the Federal Ministry of Food and Agriculture (BMEL) based on a decision of the Parliament of the Federal Republic of Germany via the Federal Office for Agriculture and Food (BLE) under the innovation support program.

\bibliography{main}

\begin{thebibliography}{16}
\expandafter\ifx\csname natexlab\endcsname\relax\def\natexlab#1{#1}\fi
\providecommand{\url}[1]{\texttt{#1}}
\providecommand{\href}[2]{#2}
\providecommand{\path}[1]{#1}
\providecommand{\DOIprefix}{doi:}
\providecommand{\ArXivprefix}{arXiv:}
\providecommand{\URLprefix}{URL: }
\providecommand{\Pubmedprefix}{pmid:}
\providecommand{\doi}[1]{\href{http://dx.doi.org/#1}{\path{#1}}}
\providecommand{\Pubmed}[1]{\href{pmid:#1}{\path{#1}}}
\providecommand{\bibinfo}[2]{#2}
\ifx\xfnm\relax \def\xfnm[#1]{\unskip,\space#1}\fi
\bibitem[{de~Berg et~al.(2008)de~Berg, Cheong, van Kreveld and
  Overmars}]{Berg2008}
\bibinfo{author}{de~Berg, M.}, \bibinfo{author}{Cheong, O.},
  \bibinfo{author}{van Kreveld, M.}, \bibinfo{author}{Overmars, M.},
  \bibinfo{year}{2008}.
\newblock \bibinfo{title}{Computational Geometry}.
\newblock \bibinfo{edition}{3rd} ed., \bibinfo{publisher}{Springer}.
\bibitem[{Chiu et~al.(2013)Chiu, Stoyan, Kendall and Mecke}]{Chiu2013}
\bibinfo{author}{Chiu, S.N.}, \bibinfo{author}{Stoyan, D.},
  \bibinfo{author}{Kendall, W.S.}, \bibinfo{author}{Mecke, J.},
  \bibinfo{year}{2013}.
\newblock \bibinfo{title}{Stochastic Geometry and Its Applications}.
\newblock \bibinfo{edition}{3rd} ed., \bibinfo{publisher}{J. Wiley \& Sons}.
\bibitem[{Colbach et~al.(2018)Colbach, Cordeau, Garrido, Granger, Laughlin,
  Ricci, Thomson and Mess{\'{e}}an}]{Colbach2018}
\bibinfo{author}{Colbach, N.}, \bibinfo{author}{Cordeau, S.},
  \bibinfo{author}{Garrido, A.}, \bibinfo{author}{Granger, S.},
  \bibinfo{author}{Laughlin, D.}, \bibinfo{author}{Ricci, B.},
  \bibinfo{author}{Thomson, F.}, \bibinfo{author}{Mess{\'{e}}an, A.},
  \bibinfo{year}{2018}.
\newblock \bibinfo{title}{Landsharing vs landsparing: How to reconcile crop
  production and biodiversity? a simulation study focusing on weed impacts}.
\newblock \bibinfo{journal}{Agriculture, Ecosystems \& Environment}
  \bibinfo{volume}{251}, \bibinfo{pages}{203--217}.
\bibitem[{Hastie et~al.(2009)Hastie, Tibshirani and Friedman}]{Hastie2009}
\bibinfo{author}{Hastie, T.}, \bibinfo{author}{Tibshirani, R.},
  \bibinfo{author}{Friedman, J.}, \bibinfo{year}{2009}.
\newblock \bibinfo{title}{The Elements of Statistical Learning}.
\newblock \bibinfo{edition}{2nd} ed., \bibinfo{publisher}{Springer}.
\bibitem[{J\"{a}hne(2005)}]{Jaehne2005}
\bibinfo{author}{J\"{a}hne, B.}, \bibinfo{year}{2005}.
\newblock \bibinfo{title}{Digital Image Processing}.
\newblock \bibinfo{publisher}{Springer}.
\bibitem[{Jungnickel(2008)}]{Jungnickel2008}
\bibinfo{author}{Jungnickel, D.}, \bibinfo{year}{2008}.
\newblock \bibinfo{title}{Graphs, Networks and Algorithms}.
\newblock \bibinfo{publisher}{Springer}.
\bibitem[{Loader(1999)}]{Loader1999}
\bibinfo{author}{Loader, C.}, \bibinfo{year}{1999}.
\newblock \bibinfo{title}{Local Regression and Likelihood}.
\newblock \bibinfo{publisher}{Springer}.
\bibitem[{Martin et~al.(2022)Martin, Lohrmann and Stoll}]{Martin2022}
\bibinfo{author}{Martin, F.}, \bibinfo{author}{Lohrmann, G.},
  \bibinfo{author}{Stoll, A.}, \bibinfo{year}{2022}.
\newblock \bibinfo{title}{Selective weed control in grassland using
  high-pressure water jets}, in: \bibinfo{editor}{{VDI Wissensforum GmbH}}
  (Ed.), \bibinfo{booktitle}{LAND.TECHNIK 2022: The Forum for Agricultural
  Engineering Innovations}, \bibinfo{address}{Düsseldorf}.
\newblock \bibinfo{note}{(submitted)}.
\bibitem[{MathWorks(2021)}]{matlab_image}
\bibinfo{author}{MathWorks}, \bibinfo{year}{2021}.
\newblock \bibinfo{title}{Image processing toolbox: Reference ({Matlab
  R2021b})}.
\newblock \URLprefix
  \url{https://mathworks.com/help/pdf_doc/images/images_ref.pdf}.
  \bibinfo{note}{(accessed 2021-11-08)}.
\bibitem[{Miettinen(2012)}]{Miettinen2012}
\bibinfo{author}{Miettinen, K.}, \bibinfo{year}{2012}.
\newblock \bibinfo{title}{Nonlinear Multiobjective Optimization}.
\newblock \bibinfo{publisher}{Springer}.
\bibitem[{Perron and Furnon(2019)}]{ortools}
\bibinfo{author}{Perron, L.}, \bibinfo{author}{Furnon, V.},
  \bibinfo{year}{2019}.
\newblock \bibinfo{title}{{OR-Tools}}.
\newblock \URLprefix \url{https://developers.google.com/optimization/}.
\bibitem[{Petrich et~al.(2020)Petrich, Lohrmann, Neumann, Martin, Frey, Stoll
  and Schmidt}]{Petrich2020}
\bibinfo{author}{Petrich, L.}, \bibinfo{author}{Lohrmann, G.},
  \bibinfo{author}{Neumann, M.}, \bibinfo{author}{Martin, F.},
  \bibinfo{author}{Frey, A.}, \bibinfo{author}{Stoll, A.},
  \bibinfo{author}{Schmidt, V.}, \bibinfo{year}{2020}.
\newblock \bibinfo{title}{Detection of colchicum autumnale in drone images,
  using a machine-learning approach}.
\newblock \bibinfo{journal}{Precision Agriculture} \bibinfo{volume}{21},
  \bibinfo{pages}{1291--1303}.
\bibitem[{Schellberg et~al.(2008)Schellberg, Hill, Gerhards, Rothmund and
  Braun}]{Schellberg2008}
\bibinfo{author}{Schellberg, J.}, \bibinfo{author}{Hill, M.J.},
  \bibinfo{author}{Gerhards, R.}, \bibinfo{author}{Rothmund, M.},
  \bibinfo{author}{Braun, M.}, \bibinfo{year}{2008}.
\newblock \bibinfo{title}{Precision agriculture on grassland: applications,
  perspectives and constraints}.
\newblock \bibinfo{journal}{European Journal of Agronomy} \bibinfo{volume}{29},
  \bibinfo{pages}{59--71}.
\bibitem[{Somerville et~al.(2020)Somerville, S{\o}nderskov, Mathiassen and
  Metcalfe}]{Somerville2020}
\bibinfo{author}{Somerville, G.J.}, \bibinfo{author}{S{\o}nderskov, M.},
  \bibinfo{author}{Mathiassen, S.K.}, \bibinfo{author}{Metcalfe, H.},
  \bibinfo{year}{2020}.
\newblock \bibinfo{title}{Spatial modelling of within-field weed populations; a
  review}.
\newblock \bibinfo{journal}{Agronomy} \bibinfo{volume}{10},
  \bibinfo{pages}{1044}.
\bibitem[{Stoll(2020)}]{Stoll2020}
\bibinfo{author}{Stoll, A.}, \bibinfo{year}{2020}.
\newblock \bibinfo{title}{{Selektive, nicht-chemische Bekämpfung von
  Giftpflanzen in extensiven Grünlandbeständen ({SELBEX})}}, in:
  \bibinfo{editor}{{Bundesanstalt für Landwirtschaft und Ernährung}} (Ed.),
  \bibinfo{booktitle}{{Tagungsband Innovationstage 20./21.Oktober 2020}},
  \bibinfo{address}{Bonn}. pp. \bibinfo{pages}{191--193}.
\bibitem[{Wegener(2020)}]{Wegener2020}
\bibinfo{author}{Wegener, J.K.}, \bibinfo{year}{2020}.
\newblock \bibinfo{title}{{Gezielter und flexibler -- Trends in der
  Pflanzenschutztechnik}}, in: \bibinfo{editor}{Frerichs, L.} (Ed.),
  \bibinfo{booktitle}{{Jahrbuch Agrartechnik 2019}}.
  \bibinfo{publisher}{{Institut f{\"u}r mobile Maschinen und Nutzfahrzeug}},
  \bibinfo{address}{Braunschweig}, pp. \bibinfo{pages}{1--7}.

\end{thebibliography}

\end{document}